\documentclass[12pt,letterpaper]{article}

\usepackage{amsmath,amssymb,array,calc,rotating,epsfig,psfrag, amscd}
\usepackage{color}
\usepackage[
      colorlinks=true,
      urlcolor=blue,    
      filecolor=blue,     
      citecolor=red,
      pdfstartview=FitV,
       bookmarksopen=true    
      ]{hyperref}
\usepackage[left=2.2cm,top=1.7cm,right=2.2cm,nohead]{geometry}

\newcommand{\ra}{\rightarrow}

\newcommand{\blp}{{\big (}}
\newcommand{\brp}{{\big )}}
\newcommand{\Blp}{{\Big (}}
\newcommand{\Brp}{{\Big )}}

\newcommand{\Bslb}{{\rm \Big [}}
\newcommand{\Bsrb}{{\rm \Big ]}}

\newcommand{\veps}{\varepsilon}

\def\lam{\lambda}
\def\om{\omega}

\def\Lam{\Lambda}

\newcommand{\bea}{\begin{eqnarray}}
\newcommand{\eea}{\end{eqnarray}}
\newcommand{\be}{\begin{equation}}
\newcommand{\ee}{\end{equation}}
\newcommand{\barr}{\begin{array}}
\newcommand{\earr}{\end{array}}

\newcommand{\cF}{{\cal F}}

\def\cO{{\cal O}}

\newcommand{\cQ}{{\cal Q}}

\newcommand{\ZZ}{{\mathbb Z}}

\newcommand{\half}{\frac{1}{2}}

\newcommand{\del}{\partial}

\newcommand{\Tr}{{\rm Tr}}

\newcommand{\tphi}{\tilde{\phi}}

\newcommand{\txi}{\tilde{\xi}}

\newcommand{\w}{\wedge}

\newcommand{\csch}{{\rm csch}}

\newcommand{\eq}[1]{(\ref{#1})}
\newcommand{\non}{\nonumber}

\newcommand{\vol}{{\rm vol}}

\newcommand{\bpm}{\begin{pmatrix}}
\newcommand{\epm}{\end{pmatrix}}
 \newcommand{\bitem}{\begin{itemize}}
 \newcommand{\eitem}{\end{itemize}}

\definecolor{cardinal}{rgb}{0.6,0,0}
\definecolor{darkgreen}{rgb}{0,0.5,0}
\definecolor{golden}{rgb}{0.92, 0.7, 0}
\definecolor{midnight}{rgb}{0, 0, 0.5}
\definecolor{darkblue}{rgb}{0.2, 0, 0.8}

\newcommand{\beq}{\begin{equation}\begin{aligned}}
\newcommand{\eeq}{\end{aligned}\end{equation}}
\newcommand{\nn}{\nonumber}

\newcommand{\fc}{\phi^c}
\newcommand{\fd}{\phi^d}

\newcommand{\fo}{\phi_0}

\newcommand{\fua}{\phi_1^a}
\newcommand{\fub}{\phi_1^b}

\def\Lam{\Lambda}

\begin{document}

\begin{flushright}
IPhT-t11/157
 \end{flushright}

\vspace{0.5cm}
\begin{center}

{\Large \bf The backreaction of anti-D3 branes on the \\
\vskip 3mm
Klebanov-Strassler geometry}\\
\vskip 3mm

 \vskip1.5cm 
 Iosif Bena$^{*}$, Gregory Giecold$^{*}$, Mariana Gra\~na$^{*}$, Nick Halmagyi$^{* \dagger}$ and Stefano Massai$^{*}$\\ 
 \vskip0.5cm
$^{*}$\textit{Institut de Physique Th\'eorique,\\
CEA Saclay, CNRS URA 2306,\\
F-91191 Gif-sur-Yvette, France}\\
\vskip0.8cm
$^{\dagger}$\textit{Laboratoire de Physique Th\'eorique et Hautes Energies,\\
Universit\'e Pierre et Marie Curie, CNRS UMR 7589, \\
F-75252 Paris Cedex 05, France}\\
\vskip0.5cm
iosif.bena, gregory.giecold, mariana.grana, stefano.massai@cea.fr\\
\vskip0.2cm
halmagyi@lpthe.jussieu.fr \\ 
\end{center}
\vskip1.5cm
\begin{abstract}

We present the full numerical solution for the 15-dimensional space of
linearized deformations of the Klebanov-Strassler background which
preserve the $SU(2) \times SU(2)\times \ZZ_2$ symmetries.
We identify within this space the solution corresponding to anti-D3
branes, (modulo the presence of a certain ``subleading'' singularity
in the infrared). All the 15 integration constants of this solution
are fixed in terms of the number of anti-D3 branes, and the solution
differs in the UV from the supersymmetric solution into which it is
supposed to decay by a mode corresponding to a rescaling of the field
theory coordinates. Deciding whether two solutions that differ in the
UV by a rescaling mode are dual to the same theory is involved even
for supersymmetric Klebanov-Strassler solutions, and we explain in
detail some of the subtleties associated to this.

\end{abstract}

\newpage

\section{Introduction}

Antibranes in warped deformed conifold Klebanov-Strassler (KS) backgrounds~\cite{Klebanov:2000hb} are a staple ingredient of string phenomenology and cosmology constructions, being essentially the only method for lifting AdS solutions with stabilized moduli, to dS solutions, and thus give rise to a landscape of dS vacua of string theory~\cite{Kachru:2003aw}.

Over the past few years we have undertaken a programme to construct the full space of first-order $SU(2) \times SU(2) \times \ZZ_2$-invariant  deformations around the KS background, in order to establish whether a solution corresponding to anti-D3 branes in this background exists, whether it has the properties one expects from the brane-probe analysis of~\cite{Kachru:2002gs}, and whether it is dual to a metastable vacuum of the dual boundary theory. The underlying philosophy of this programme has been that one cannot decide a-priori that a metastable anti-D3 brane solution must exist, and then accept whatever boundary conditions are necessary in order for this to happen, but rather one should start from a set of physical infrared and ultraviolet boundary conditions, and ask whether a solution compatible with these boundary conditions exists or not.

The key results of this investigation have been: \vspace{.1cm}\\
1.~One can find all the homogeneous solutions to, and thus solve implicitly the equations~\cite{Borokhov:2002fm} governing the first-order perturbations. The full solution seems at first to involve 8 nested integrals~\cite{Bena:2009xk}. \vspace{.1cm}\\
2.~One can simplify these and write the full solution in terms of 2 nested integrals~\cite{Bena:2011hz}, which are in fact integrals of rational functions multiplying the warp factor and Green's function of the KS background. \vspace{.1cm}\\
3.~One can write the UV and IR expansions of the generic solution to this space of deformations, and identify all the UV normalizable and non-normalizable modes, as well as the infrared physical boundary conditions for D-branes~\cite{Bena:2009xk}. \vspace{.1cm}\\
4.~The force on a probe D3 brane in the first-order perturbed background depends only on {\it one} of the 16 integration constants, and this constant must be nonzero if the solution is to correspond to antibranes~\cite{Bena:2009xk}. Furthermore, the full functional expression of this force can be calculated~\cite{Bena:2010ze}, and matches exactly the expression one obtains from ``Newton's third law'' arguments \`a la KKLMMT~\cite{Kachru:2003sx}. 
 \vspace{.1cm}\\
5. The putative solution for anti-D3 branes smeared on the three-sphere at the tip of the KS solution is expected to have a singularity in the five-form and warp factor, coming from the physical brane sources. Besides this, the solution must also have a subleading singularity, proportional to the coefficient of the brane-attracting mode of the solution.

As explained in \cite{Bena:2009xk}, if the singularity is not physical, then the backreaction of anti-D3 branes in the KS solution gives rise to a large deformation of this solution, which cannot be captured in perturbation theory, much like when one tries to construct metastable vacua using type IIA brane engineering~\cite{Bena:2006rg}. On the other hand, if the singularity is physical, then our technology produces the full first-order backreacted solution corresponding to antibranes in the KS background, as well as all first-order deformation of the KS solution by non-normalizable $SU(2) \times SU(2) \times \ZZ_2$-invariant modes, corresponding to all the relevant and irrelevant deformations of the dual field theory.

This subleading singularity cannot be attributed to any brane source (it has the wrong orientation), or to brane-flux annihilation (it is linear in the antibrane number, while the brane-flux annihilation is nonlinear). However, as mentioned in~\cite{Bena:2009xk} and argued in~\cite{Dymarsky:2011pm}, it is possible that this singularity is an artifact of perturbation theory, and may not be present in a fully-backreacted solution for antibranes. On the other hand, obtaining a fully-backreacted solution for antibranes in ISD flux backgrounds seems to run into trouble in less complicated setups \cite{Blaback:2010sj, Blaback:2011nz}, and can even be ruled out by topological arguments (that yield a physics similar to the one found in \cite{Bena:2006rg}). If the results of
\cite{Blaback:2011nz} extend to the KS solution, then the presence of a subleading singularity in perturbation theory will look with hindsight as an indication of a more profound problem with the whole construction.

Given that the arguments about this singularity fall mainly outside of the scope of our perturbation theory machinery, it is best to hedge our bets both ways, and ask whether inside the 15-dimensional space of parameters that characterize our first-order solution one can identify a solution that has the correct physics to correspond to anti-D3 branes in the KS geometry, subleading singularity aside. Identifying this solution inside the 15-dimensional space is simpler than finding a needle in a haystack, but not by far: One has to throw away divergent terms both in the UV and in the IR expansion~\cite{Bena:2009xk}, and to impose the correct D-brane boundary conditions on the divergence of the warp factor and electric field in the infrared. 

Those conditions yield algebraic relations between the various integration constants that appear in the UV or IR expansions of the fields; however, the integration constant that appears in the UV expansion of a given field, say the dilaton, is not the same as the one that appears in its IR expansion, but differs by highly nontrivial combination of the other integration constants. Hence, even if we impose all the physical boundary conditions in the UV and in the IR, we are far from being done, because the UV conditions are expressed using the UV integration constants, and the IR conditions are expressed using the IR integration constants, and it is possible that upon translating the UV conditions into IR variables one may have the unpleasant surprise that these conditions are incompatible. Hence, in order to identify the correct antibrane solution inside the 15-dimensional space of first-order deformations, and to establish whether this solution is dual or not to a metastable vacuum of a supersymmetric field theory, it is crucial to relate the UV and IR expansion coefficients, which is the main purpose of this paper.

Before unveiling those results, we would like to point out that identifying whether two asymptotically-KS supergravity solutions are dual to vacua of the same field theory is not as straightforward as it might seem, even for supersymmetric solutions, essentially because, besides the seven normalizable and seven non-normalizable deformations, there exists another deformation corresponding to rescaling the field theory directions. Of course, if two solutions differ by non-normalizable deformations, they clearly are dual to two different field theories; however, as we will explain in Section~\ref{section-distinguish}, two solutions with different rescaling parameters may or may not belong to the same theory. Hence, using purely UV data one cannot distinguish asymptotically-KS supersymmetric solutions that we expect~\cite{Dymarsky:2005xt} to be dual to different field theories, unless one introduces extra assumptions about the infrared of the solutions, or about their bulk behavior.

Anticipating our results, we compute the unique solution that has the correct infrared and ultraviolet divergences (modulo the subleading singularity) to describe anti-D3 branes in the KS background. All the parameters of this solutions can be determined in terms of the number of antibranes. Nevertheless, much like for supersymmetric solutions, one cannot distinguish using purely UV data whether this solution describes a non-supersymmetric vacuum of a supersymmetric solution, or whether it is dual to a distinct non-supersymmetric theory. To achieve that one must therefore introduce extra assumptions about the infrared or about the bulk.

In Section~\ref{deformations} we give a lightning review of the general construction of first-order deformations around the KS solution (the full details can be found in Appendix~\ref{app:BG}), and in Section~\ref{oldsolution} we 
review the simplified analytic solution found in~\cite{Bena:2011hz} in terms of two nested integrals (whose full details can be found in Appendix~\ref{app:solution}).  In Section~\ref{relating} we explain the procedure we use to relate the UV and the IR integration constants, and illustrate with more details how this procedure can be implemented for one of the perturbation modes. We also give the relations between the UV and IR integration constants of the other modes; the derivation of all these relations is left for Appendix~\ref{app:IRUVexps}. In Section~\ref{section-distinguish} we present the different criteria for distinguishing supersymmetric asymptotically-KS solutions, and in Section~\ref{antiD3solution} we identify the solution for anti-D3 branes inside the space of solutions. Section~\ref{oddsandends} is devoted to the relation between our solution and the one obtained in~\cite{DeWolfe:2008zy} by perturbing around the Klebanov-Tseytlin (KT) solution, and to the identification within our space of solutions to perturbation of the KS solution by non-normalizable modes dual to gaugino masses.

\section{Non-supersymmetric deformations around the \\ Klebanov-Strassler background}
\label{deformations}

\subsection{Ansatz and background solution}

We wish to construct the backreacted solution corresponding to $\bar N$ anti-D3 branes smeared on the $S^3$ at
the tip of the warped deformed conifold. We use the Ansatz proposed by Papadopoulos and Tseytlin~\cite{Papadopoulos:2000gj}, which is the most general one (with vanishing RR axion $C_0$) that preserves the $SU(2)\times SU(2)\times \ZZ_2$-symmetry of the Klebanov-Strassler solution (KS). The metric is 
\beq\label{PTmetric}
ds_{10}^2 = e^{2\, A+2\, p-x}\, ds_{1,3}^2 + e^{-6\, p-x}\, d\tau^2 + e^{x+y}\, \left( g_1^2 + g_2^2 \right) + e^{x-y}\, \left( g_3^2 + g_4^2 \right) + e^{-6\, p-x}\, g_5^2 \ ,
\eeq
where all the functions depend on the radial variable $\tau$. The fluxes and the dilaton are
\bea\label{PTfluxes}
 H_3 &=& \tfrac12 \, \left( k - f \right)\, g_5 \wedge \left( g_1 \wedge g_3+ g_2 \wedge g_4 \right) + d\tau \wedge \left( f'\, g_1 \wedge g_2+ k'\, g_3 \wedge g_4 \right) \ , \nn \\
 F_3&=& F\, g_1 \wedge g_2 \wedge g_5 + \left( 2\, P - F \right) \, g_3\wedge g_4 \wedge g_5 +
 F' \, d\tau \wedge \left( g_1 \wedge g_3 + g_2 \wedge g_4 \right) \ , \\
 F_5 &= & {\cal F}_5 + * {\cal F}_5 \, , \qquad {\cal F}_5 = \left[\frac{\pi \, Q}{4}+ (k-f)\, F +  2\, P\, f  \right] \, g_1 \wedge g_2 \wedge g_3 \wedge g_4 \wedge g_5  \ ,\nn \\
 \Phi&=& \Phi(\tau)\, , \qquad C_0 = 0  \nn \, ,
 \eea
where $P$, $Q$ are constants while $f,k$ and $F$ are functions of $
\tau$. A prime denotes a derivative with respect to $\tau$.

The fields from this Ansatz are collectively denoted $\phi^a$, $a=1,...,8$.  We will study and fully determine the solution space of first-order non-supersymmetric deformations of the supersymmetric Klebanov-Strassler theory,
\beq
\phi^a = \phi^a_0 + \phi^a_1(Z) + {\cal O}(Z^2)\, .
\eeq

The background fields $\phi^a_0$ are given by the Klebanov-Strassler solution without mobile D3-branes:
 \bea \label{KSbackground}
 e^{x_0}&=& \frac14 \, h(\tau)^{1/2} \, \left( \tfrac12 \, \sinh(2\, \tau) - \tau \right)^{1/3} \, , \non \\
 e^{y_0}&=&\tanh(\tau/2) \, , \non \\
 e^{6\, p_0}&=&  24\, \frac{\left( \tfrac12 \, \sinh(2 \, \tau) - \tau \right)^{1/3}}{ h(\tau) \, \sinh^2\tau}  \, , \non \\
 e^{6\, A_0}&=&\frac{\veps_0^4}{3 \cdot 2^9} \, h(\tau)\, \left( \tfrac12 \, \sinh(2 \, \tau) - \tau \right)^{2/3}\, \sinh^2\tau \, , \\
 f_0&=& - P \, \frac{\left( \tau \, \coth \tau -1 \right)\, \left( \cosh \tau -1 \right)}{\sinh \tau} \, , \non\\
 k_0&=&- P\, \frac{\left(\tau \, \coth \tau -1 \right)\, \left(\cosh \tau +1 \right)}{\sinh \tau} \, , \non \\
 F_0&=& P\, \frac{\left(\sinh \tau -\tau \right)}{\sinh \tau} \, , \non \\
 \Phi_0&=&0 \, , \non  \\
 Q&=&0 \ , \nn
 \eea
where $\veps_0$ is the deformation parameter of the conifold, related to the confinement scale of the dual gauge theory. Of significance are also the warp factor $h$ and the Green's function $j$ for this background: 
\begin{align}
h(\tau) &= 32\, P^2\, \int_\tau^{\infty} \frac{u\, \coth u - 1 }{\sinh^2 u}\, \left( \cosh u \, \sinh u-  u \right)^{1/3}\, du \, , \label{hintegral} \\
j(\tau)&
= -\int_{\tau}^{\infty} \frac{du}{\left( \cosh u \, \sinh u - u \right)^{2/3}}\, .  \label{Gintegral}
\end{align} 

Note that the last equality in~\eqref{KSbackground} implies we are taking $g_s=1$. Furthermore, the dimensionful constant $P$ is related to the quantized dimensionless units of flux $M$ entering in the rank of the gauge groups of the dual field theory (see section \ref{sec:dictionary}) by
\beq \label{PM}
P=\frac14 \, M\, \alpha' \ ,
\eeq
So as to avoid extra clutter, in what follows we take $\alpha'=1$, and $\veps_0=1$.

\subsection{First-order perturbation equations, conditions and physical significance of the integration constants}

Using a method due to Borokhov and Gubser~\cite{Borokhov:2002fm} and reviewed in the Appendix~\ref{app:BG}, finding linearized deformations away from a supersymmetric solution, can be reduced to solving two sets of first-order ordinary differential equations in the radial variable $\tau$, instead of second-order differential equations. Out of those two sets, the first one forms a closed system for the variables $\xi_a$ that can be thought of as ``conjugate momenta" for the perturbations $\phi_1^a$ of the fields entering our Ansatz~\eqref{PTmetric},~\eqref{PTfluxes}. The integration constants associated to that first system are labelled $X_a$, and are non-zero for a non-supersymmetric solution. The integrations constants from the second system of coupled 1st-order ODE's are denoted $Y_a$. 

For the problem of present interest, i.e.~the backreaction of anti-D3's on KS, the solution to the system was found in~\cite{Bena:2009xk,Bena:2011hz} after applying the following change of basis  \footnote{\label{foot:thpi4}Note that $\tilde{\phi}_4$ is the perturbation to the warp factor, namely $\tilde{\phi}_4= - 2 \, \tilde{A}$, since the warp factor of the KS theory~\eqref{hintegral} is such that $h(\tau) \equiv e^{4\, \tilde{A}} = e^{4\, A + 4\, p - 2\, x}$. Cf. also equation~\eqref{tilde A GKP} below.} 
 \be
 \tphi_a= \left( x - 2\, p - 5\, A, \, y, \, x+3\, p, \, x-2\, p - 2\,A, \, f + \frac{\pi \, Q}{8\, P}, \, k+\frac{\pi \, Q}{8\, P}, \, F, \, \Phi \right) \ , \label{tphidef}
 \ee

There is one relation between the constants $X_a$ that has to be obeyed on the whole space of solutions. Namely, the zero-energy condition
\be \label{ZEC}
6\, X_2 - 4\, X_3 - 6\, P\, X_5 - 9\, P\, X_7 = 0 \ .
\ee   
Another integration constant, $Y_1$ as it happens, looks naively like it can be gauged away by a rescaling of the four-dimensional coordinates but as we will see later plays a crucial role in the physics. We are therefore left with fifteen meaningful integration constants.  

Out of those fifteen parameters, the one called $X_1$ plays a key role. Indeed, the force exterted on a probe D3-brane is directly proportional to it and does not depend on any other integration constant~\cite{Bena:2009xk}. Its expression was found in~\cite{Bena:2010ze} and is given by
\bea
F_{D3_+}&=&\frac{2}{3}\, e^{-2x_0} \xi_1 \nn\\
&=&\frac{2}{3}\, e^{-2x_0}\, X_1\, h(\tau) \, , \non \\
&=&\frac{32}{3}\, \frac{2^{2/3}\, X_1}{(\sinh 2\, \tau - 2\, \tau)^{2/3}}\label{probeD3force} \, .
\eea
One can also use the conventions of~\cite{Giddings:2001yu} to describe the same result for a first order expansion around any warped Calabi-Yau background with ISD flux. Here the derivative of the DBI and WZ actions for D3-branes are respectively proportional to the warp factor $e^{4\, \tilde{A}}$ and the four-form RR potential $C_4=\alpha \, dx^0 \wedge ... \wedge dx^3$, where 
in the language of~\eqref{PTmetric} and~\eqref{PTfluxes}, we have
\beq\label{tilde A GKP}
\tilde A=A+p-\frac{x}{2} \ , \qquad \alpha'= - e^{4\, A + 4\, p - 4\, x}\, \left[\frac{\pi \, Q}{4} + k\, F + f\, \left( 2\, P - F \right) \right] \ .
\eeq
The force is found to be
\beq \label{Phipm}
F_{D3_\pm} = \Phi_\mp ' \ , \quad \text{where} \ \Phi_\pm=e^{4\, \tilde A} \pm \alpha \ ,
\eeq  
and by D3$_-$ we mean $\overline{D3}$-branes. The combinations $\Phi_\pm$ are sourced by D3$_{\pm}$ respectively, and by $|G_\pm|^2$~\cite{DeWolfe:2004qx,Dymarsky:2011pm} where $G_\pm=G_3 \mp i *G_3$ and $G_3=F_3+i e^{-\phi}H_3$. 

\section{Our analytic solution for the full space of first-order deformations around KS}
\label{oldsolution}

In a previous work~\cite{Bena:2011hz}, we found that the fully analytic generic solution to the most general first-order deformation of the Klebanov-Strassler background involves at most two nested integrals of the form
\be
\int^{\tau} h(u)\, f(u)\, du \ , \quad \text{or} \quad  \int^{\tau} j(u)\, f(u)\, du\,,
\ee
where $f(\tau)$ is a certain combination of hyperbolic functions. Expressions for the warp factor $h(\tau)$ of the KS background and its Green's function $j(\tau)$ are provided in~\eqref{hintegral} and~\eqref{Gintegral}.

\begin{figure}[h]
\begin{center}
\includegraphics[scale=1]{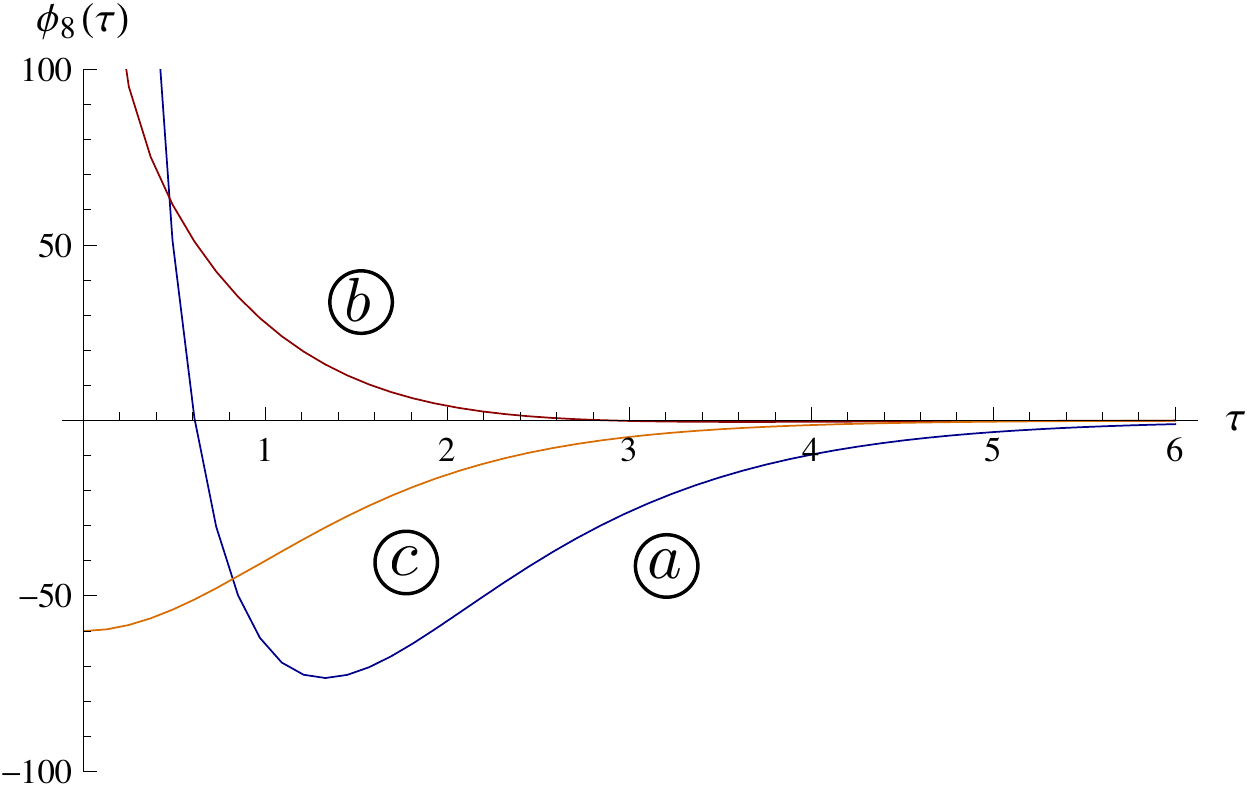} 
\caption{The profile of the field $\tphi_8$ corresponding to a shift
  of the dilaton, for the following choices of integration constants (with e.g.~$P=1$). Blue, also labelled $\textcircled{a}$: $X_1 = 1, X_5 = -\frac{15}{2}, X_6=X_7=5, X_8=2,
  Y_8=-88.05$; Red $\textcircled{b}$: $X_1 = X_6=X_7=1, X_5 = -\frac{7}{6}, X_8=1.8,
  Y_8=-111.5$; Yellow $\textcircled{c}$: $X_1 = X_7=2, X_5 = -\frac{7}{6}, X_6=8.608, X_8=-0.843,
  Y_8=-133.9$. In each case, $Y_8$ is fixed so as to ensure that $\tilde \phi_8(\infty)=0$.} \label{GraphicsPhi8numonly}
\end{center}
\end{figure}

Let us illustrate this with the result for $\tphi_8$, corresponding to shifts in the dilaton. The analytic solutions for all seven remaining modes are consigned to Appendix~\ref{app:solution}, and more details of the derivation can be found in \cite{Bena:2009xk,Bena:2011hz}.
\begin{align}\label{phi8}
\tphi_8 =&\, Y_8 - 64\, X_8\, j(\tau) + \frac{X_7}{P}\, h(\tau) \non\\ & - 64\, P\, X_6\, \int_1^{\tau} \frac{\left( u\, \coth u - 1 \right)}{\sinh^2 u \, \left( \cosh u \, \sinh u - u \right)^{2/3}} \, d u \non\\ & + 2 \, \frac{X_5}{P}\, h(\tau)  + \frac{16}{3}\, X_1\, \text{csch}^2 \tau \, \left( \cosh \tau \, \sinh \tau - \tau \right)^{1/3}\, h(\tau) \non\\ & + \frac{64}{9}\, X_1\, h(\tau) \, j(\tau) 
- \frac{32}{9}\, X_1\, \int_1^{\tau} \frac{\left( \sinh^2 u + 1 - u \, \coth u \right)}{\sinh^2 u \, \left( \cosh u \, \sinh u - u \right)^{2/3}}\, h(u)\, d u \, .
\end{align}
We have chosen to integrate in the domain $[1,\tau]$, given that many of the integrands 
(like the one from the last term above) are infrared-divergent.
Once the limits of integration are fixed, the constant $Y_8$ in~\eqref{phi8} is defined unambiguously. The profile for $\tphi_8$ is given in Figure~\ref{GraphicsPhi8numonly}.

The infrared and ultraviolet behaviors of the modes are given in Appendix~\ref{app:IRUVexps}. Some of the integration constants appearing in the infrared expansions (like  $Y_3^{IR}$ or $Y_6^{IR}$) correspond to unphysical divergences of various fields, and we will set them to zero. Other constants (like $Y_7^{IR}$ or $X_1$) correspond to physical divergences in the warp factor and in the RR five-form field strength coming from the presence of smeared anti-D3 branes, and we need to keep them in the final solution. We will explain this procedure  when we construct the antibrane solution in section \ref{antiD3solution}.

In order to stress out how the integration constants $X_a$ and $Y^{a}$ are paired into normalizable and non--normalizable modes we also remind the reader of the UV behaviors of those modes \cite{Bena:2009xk}, which one can also extract from the expansions in Appendix~\ref{app:IRUVexps}:
 \begin{table}[h]
\label{UVmodetable}
\begin{center}
\begin{tabular}{|c|c|c|c|c|c} \hline
dim $\Delta$ & non-norm/norm & integration constants \\ \hline
8 & $r^4/r^{-8}$ & $Y_4/X_1$  \\\hline
7 & $r^3/r^{-7}$ & $Y_5/X_6$ \\\hline
6 & $r^2/r^{-6}$ & $X_3/Y_3$ \\\hline
5 & $r/r^{-5}$ & $---$ \\\hline
4 & $r^0/r^{-4}$ & $Y_7,Y_8,Y_1/X_5,X_4,X_8$ \\\hline
3 & $r^{-1}/r^{-3}$ & $X_2,X_7/Y_6,Y_2$ \\\hline
2 & $r^{-2}/r^{-2}$ & $---$ \\\hline
\end{tabular}
\caption{The UV behavior of all sixteen modes for the $SU(2)\times SU(2)\times \ZZ_2$-symmetric deformation Ansatz around the Klebanov-Strassler solution.}
\end{center}
\end{table}


\section{Relating the IR and UV integration constants}\label{subsecIRUV}
\label{relating}

Given that ultimately we will have to impose boundary conditions on the generic analytic
solution to the full space of first order deformations around KS, we should look at the IR and UV behavior of
the modes $\tphi_a$. Their somewhat lengthy analytic expressions are gathered in Appendix~\ref{app:solution} and were first found in~\cite{Bena:2011hz}. Moreover, it is not enough to consider the
expansions shown in Appendix~\ref{app:IRUVexps}. The zeroth-order
terms in the expansions collected in that Appendix include arbitrary integration constants
coming from indefinite integrations, which are generically denoted as
$Y_a^{IR}, Y_a^{UV}$. In order to determine how the $Y_a^{IR}$'s are related to the $Y_a^{UV}$'s and
thus to connect the IR and UV regions, we have to perform a numerical
integration that will fix $Y_a^{UV}$ as follows:
\begin{equation}
Y_a^{UV} = Y_a^{IR} + \sum_{b=1}^{8} \mathbf{N}_a{}^b\, X_b \, ,
\end{equation}
where $\mathbf{N}$ is a matrix of numerical coefficients arising out of
evalutions of the single and double integrals appearing in the analytic solutions for the $\tphi_a$ modes.

\subsection{Our results}

All in all, following the procedure we have just outlined, the relations between all\footnote{Except $Y_4$, which is far more difficult to get and will not be needed for our following analysis in any case.}  the $Y_a^{UV}$ and $Y_a^{IR}$ that we have derived are as follows:

\begin{equation}\label{YUVYIR}\renewcommand\arraystretch{1.2}
\begin{pmatrix} Y_8^{UV} \\ Y_2^{UV} \\  Y_3^{UV}\\Y_1^{UV} \\ Y_5^{UV} \\
  Y_6^{UV}\\Y_7^{UV}\end{pmatrix}  = \begin{pmatrix} Y_8^{IR} \\
  Y_2^{IR} \\ Y_3^{IR} - 2\, Y_2^{IR} \\Y_1^{IR}-\frac53\, Y_2^{IR}\\ Y_5^{IR} + \frac{P}{6}\, Y_8^{IR}\\
  Y_6^{IR}+ \frac{3\, P}{2}\, Y_2^{IR} - \frac{P}{2}\, Y_8^{IR}\\ Y_7^{IR} - P\,
  Y_2^{IR} + P\, Y_8^{IR}\end{pmatrix}  + \mathbf{N} \cdot  \begin{pmatrix} X_1\\ X_2\\ X_3\\ X_4\\
  X_5\\X_6\\ X_7 \\X_8 \end{pmatrix} ,
\end{equation}
with the matrix $\mathbf{N}$
\begin{align}
\mathbf{N} &= \nn \\& 
\renewcommand\arraystretch{1.2}\begin{pmatrix}
 -235.3\, P^2 & 0         & 0         & 0         &- 36.47\, P   & 35.71\, P   & -18.24\, P   &  53.56 \\
-3.870\, P^2  & 0         & 83.34  & 7.791  & 83.34\, P     & -12.37\, P  &   166.7\, P & 0\\
93.63\, P^2  & 250.0   &206.7   & 93.84  & -284.0 P & 61.22\, P &-243.8\, P & 0\\
-123.8\, P^2 & -40.25  &  70.31 & -1.827 & 22.93\, P & 35.50\, P    & 71.33\, P &0\\
-165.9\, P^3 &-20.16\, P & 19.52\, P& 1.488\, P&14.08\, P^2 & 11.90\, P^2 &36.32\, P^2 & 17.85\, P \\
100.6\, P^3   &-166.7\, P & 81.27\, P &-46.06\, P&221.4\, P^2&-48.57\, P^2&265.8\, P^2&-8.545\, P\\
-225.8\, P^3  &83.34\, P  &-94.65\, P &16.52\, P&-158.9\, P^2 &35.92\, P^2&-221.4\, P^2&17.09\, P 
\end{pmatrix}. \nn
\end{align}

The above relations~\eqref{YUVYIR} depend at an intermediary stage on our results for the relation between the integration constants $Y_a$ that
appear in the analytic solution~\eqref{phi8appendix}--\eqref{applambda7} and the constants
$Y_a^{IR}$ that appear in the IR expansions~\eqref{phi8IR}--\eqref{phi4IR}, obtained via the method
summarized at the beginning of this section and further expanded upon in the next subsection. We provide them here as a matter of having accessible intermediate results:
\begin{equation}\label{appYIRtoY}\renewcommand\arraystretch{1.2}
\begin{pmatrix} Y_8^{IR} \\ Y_2^{IR} \\  Y_3^{IR}\\Y_1^{IR} \\ Y_5^{IR} \\
  Y_6^{IR}\\Y_7^{IR}\end{pmatrix}  =\begin{pmatrix}
 Y_8 \\
Y_2 \\
 2\, Y_2+ Y_3 \\
 Y_1 \\
 Y_5 - \frac{P}{6}\, Y_8 \\
 -\frac{P}{2}\, Y_2 + Y_6 \\
 P\, Y_2 + Y_7
\end{pmatrix} + \mathbf{M}_{(Y^{IR},Y)}\cdot  \begin{pmatrix} X_1\\ X_2\\ X_3\\ X_4\\
  X_5\\X_6\\ X_7 \\X_8 \end{pmatrix} ,
\end{equation}
\begin{align}
\mathbf{M}_{(Y^{IR},Y)}&= \nn \\& \hspace{-0.8cm}
\renewcommand\arraystretch{1.2}\begin{pmatrix}
 352.6\, P^2 & 0 & 0 & 0 & 36.47\, P & -41.56\, P & 18.24\, P & -53.56 \\
 25.86\, P^2 & 0 & -33.23 & 3.918 & -38.81\, P & -3.432\, P & -69.25\, P & 0 \\
 -18.62\, P^2 & -99.69 & 15.54 & -0.9673 & 7.797\, P & -7.959\, P & 15.92\, P & 0 \\
 144.4\, P^2 & 98.79 & -67.47 & 5.146 & -81.34\, P & -44.35\, P & -153.9\, P & 0 \\
 92.62\, P^3 & 12.26\, P & -9.501\, P & -4.435\, P & -16.54\, P^2 & -18.03\, P^2 & -22.52\, P^2 & -11.85\, P \\
 8.129\, P^3 & 24.44\, P & -1.632\, P & 1.147\, P & -4.773\, P^2 & 2.180\, P^2 & -11.20\, P^2 & -3.979\, P \\
 -1.307\, P^3 & -38.81\, P & -4.754\, P & 3.491\, P & 1.749\, P^2 & 3.599\, P^2 & -6.256\, P^2 & 7.959\, P
\end{pmatrix}.\nn
\end{align}
Analogously, the link between the parameters $Y_a^{UV}$ and $Y_a$ can similarly be obtained from the UV/IR relation~\eqref{YUVYIR}. 

\subsection{An illustration of the procedure}

As an example making this procedure plainer to the reader, we show how we relate $Y_8^{UV}$ and $Y_8^{IR}$. This is a three-stage procedure:\\
(i) first, we relate $Y_8^{IR}$ and the parameter $Y_8$ appearing in~\eqref{phi8};\\
(ii) we next obtain the relation between $Y_8^{UV}$ and $Y_8$;\\
(iii) finally, using results from the above steps, we get $Y_8^{UV}$ in terms of $Y_8^{IR}$.  
 
In order to implement step (i) above and relate $Y_8^{IR}$ to $Y_8$, we expand the integrands entering the IR expansion of the solution to the $\tphi_8$ equation up to a certain power in $\tau$. We then evaluate the indefinite integral and call $Y_8^{IR}$ the constant term in $\tphi_8$. The first few terms
in those expansions are given by~\eqref{phi8IR}, which we provide here for convenience:
\bea \label{phi8IRhere}
\tilde \phi_8^{IR}&=&\frac{1}{\tau}\Big( \frac{32}{3}
\left(\frac{2}{3}\right)^{1/3} \, ( 3PX_6 -h_0 X_1 ) + 32\cdot
2^{1/3}\cdot3^{2/3} \, X_8 \Big) + Y_8^{IR} +\cO(\tau) \ .
\eea

\begin{figure}[h]
\begin{center}
\includegraphics[scale=0.85]{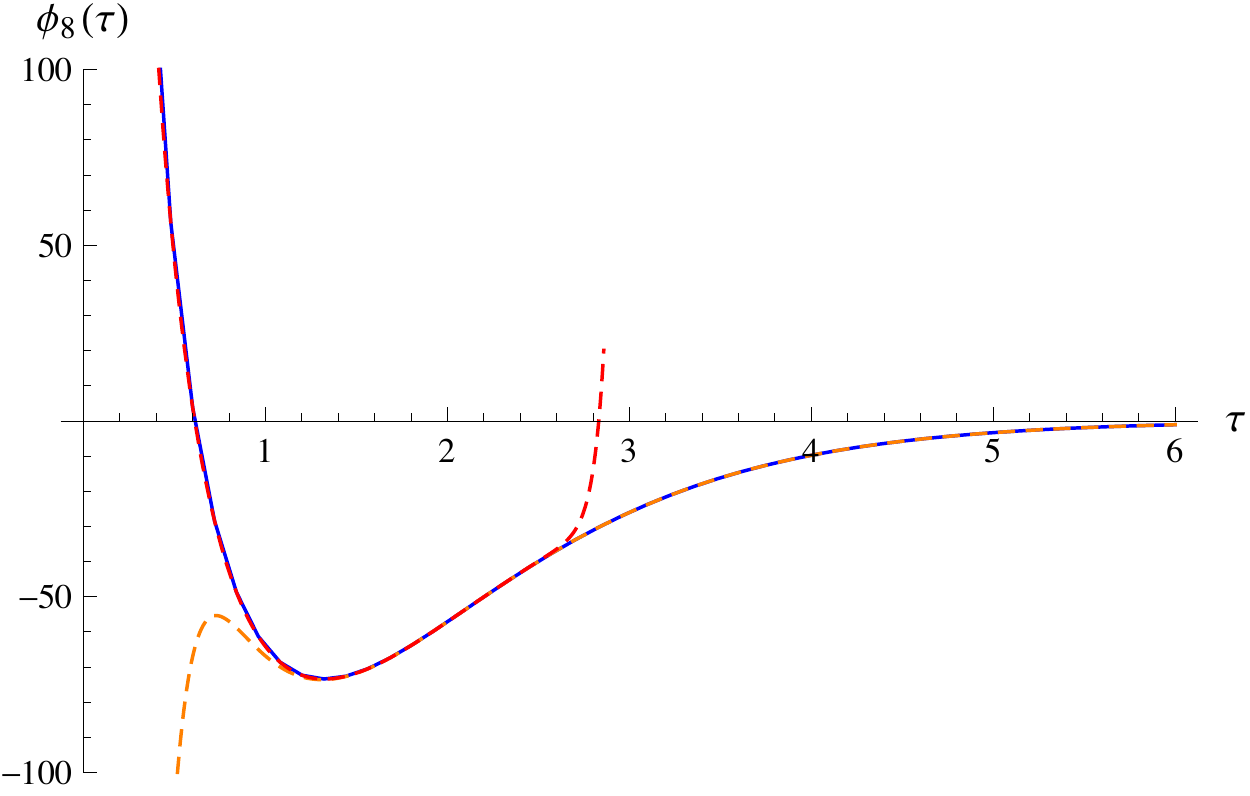} 
\caption{The numerical solution for the field $\tphi_8$ for $X_1 = 1, X_5 = -\frac{15}{2}, X_6=5, X_7=5, X_8=2,
  Y_8=-88.05, P=1$ (underlying blue solid
  line). The red and orange dashed lines correspond respectively to the IR and UV expansions.} \label{GraphicsPhi8}
\end{center}
\end{figure}

We now have to match~\eqref{phi8IRhere} at some small $\tau$ with the
numerical value of $\tphi_8$ that we obtain by performing the integrals
in~\eqref{phi8} numerically. 
Since the expansions for the integrands are good up to $\tau>1$, we did choose to match at $\tau=1$, where the integrals that enter the solutions for the $\tphi$'s are zero by definition. Evaluating numerically~\eqref{phi8} at $\tau = 1$, we find
\beq\label{xpz1}
\tilde \phi_8(\tau=1)=Y_8 + 84.0493\, P^2 \, X_1 + 28.5159\, P\, X_5 + 14.2579 \, P
\, X_7 + 41.2221\, X_8 \, ,
\eeq
while from the IR expansion of $\tphi_8$~\eqref{phi8IRhere}, we have
\bea\label{xpz2}
\tilde \phi_8^{IR}(1)&=&Y_8^{IR}-268.524 \, P^2 \, X_1-7.9588\,  P\,  X_5+
41.5621\,  P\, X_6 \nn\\
&&-3.97940 \, P\, X_7+ 94.786\, X_8 \, .
\eea
Comparing the above two results,~\eqref{xpz1} and~\eqref{xpz2}, we finally obtain the end-result of step (i) above:
\bea
Y_8^{IR}&=&Y_8 + 352.574\, P^2\, X_1 + 36.4747\, P \, X_5 - 41.5621\, P\, X_6  \nn \\
&&+   18.2373\, P\, X_7 - 
  53.5642\, X_8 \, .
\eea
With this relation at hand, we can furthermore make sure, as one more consistency test, that the numerical integrals and the series agree at small $\tau$. The result is shown on Figure~\ref{GraphicsPhi8}. 

We go through the same recipe for the UV and compare the value of the UV series of
the integrands with the value of $\tphi_8$ that we have obtained by
performing the integrals numerically\footnote{With as much precision as desired. Here, for both IR and UV expansions, we have settled for 20 orders of WorkingPrecision using Mathematica. The UV series expansions were derived up to order 15.} at $\tau = 15$.
When the dust settles down, we find the following relation between $Y_8^{UV}$ and $Y_8$:
\beq
Y_8^{UV}= Y_8 + 117.318\, P^2 \, X_1 - 5.85263\, P\, X_6 \, . 
\eeq

As one extra check, inserting the above result in the UV expansions, we can verify that the UV series approximates well our numerical results at large $\tau$. This can also be see on Figure~\ref{GraphicsPhi8}.

Note that for $\tphi_8$ there is a rather large range of overlap between its IR and UV series expansions. So, with hindsight, for this particular mode, we could have avoided going through tedious numerical work. On the other hand, for most of the other $\tphi^{a}$ fields, the overlap is much narrower. Therefore, in order to attain satisfactory precision in relating the IR and UV integration constants, we have opted for a careful numerical analysis.


\section{Asymptotically KS solutions and their field theory interpretation}
\label{section-distinguish}

Having found the full 15-dimensional space of perturbative solutions around the KS background, we would now like to develop the machinery that will allow us to identify whether the antibrane solution is in the same theory 
as the supersymmetric background into which it is conjectured to decay \cite{Kachru:2002gs}. However, as mentioned in the introduction, distinguishing between asymptotically-KS solutions and arguing which background is dual to which field theory using only UV data is not trivial even for supersymmetric solutions, essentially because of the existence of the scale deformation $Y_1$, which equivalently can be traded for the $\veps$ parameter that characterizes the size of the deformed conifold before the warping.

If two solutions differ by non-normalizable deformations, they are dual to two different field theories. However, our fifteen-dimensional deformation space has the peculiarity that there are seven pairs of normalizable/non-normalizable modes and then one extra mode $Y_1$. The putative partner to $Y_1$ is eliminated by the zero-energy condition and it may seem that $Y_1$ itself is a gauge artifact which can be removed by rescaling the four-dimensional space-time coordinates. As we will mention in more detail below, while for a single vacuum this is true, if there are two isolated vacua in the same theory then there remains a dimensionless number (essentially the ratio of the confinement scales) which can be attributed to $Y_1$.

One can inquire whether two solutions that have the same non-normalizable modes but two different $\veps$'s, hence two different scale deformations, are dual to the same field theory. The answer is not clear, because one can change $\veps$ and at the same time change also the number of mobile branes, keeping the total charge at infinity constant. Changing $\veps$ changes the volume of the space, and since the space has charge dissolved in flux, one also changes the total charge; one can compensate for this change by introducing or taking away mobile branes.

Hence, a vacuum with no mobile branes and one value for $\veps$ has exactly the same UV data as a vacuum with one mobile brane and another value of $\veps$, or a vacuum with, say, 17 mobile branes and yet another value of $\veps$. Clearly these solutions cannot be all dual to vacua of the same KS field theory. On the other hand, a background with $M$ mobile branes (where $M$ is the amount of RR three-form flux on the KS three-cycle) and a certain value of $\veps$ and another one with no mobile branes were argued in~\cite{Dymarsky:2005xt} to be dual respectively to the mesonic vacuum and the baryonic vacuum of the same $SU(k M) \times SU(kM+M)$ theory. Hence, even in the supersymmetric theory, one cannot decide whether two vacua with different scales and different amounts of mobile branes are in the same theory by simply examining their UV data.

In this section, we discuss the supersymmetric KS situation in detail, and argue that in order to be able to use UV data to distinguish between two supersymmetric asymptotically-KS solutions that should {\it not} be dual to the same theory, one must introduce an additional criterion. The most obvious choice is requiring that the value of the NSNS $B_2$ field that wraps the $S^2$ which shrinks to zero size at the conifold tip must be zero, and can only jump by integral periods. 
%
%
After all, the $S^2$ is topologically trivial, and if the integral of $B_2$ is nonzero, one can stay at a fixed radius, consider a very small closed fundamental string at the north pole and take it around the $S^2$ to the south pole; during this process its world-sheet action will pick up a phase proportional to the $B_2$ integral. If one now brings back the string to the north pole, the string will interfere destructively with itself unless the integral of $B_2$ on $S^2$ is an integer\footnote{We thank Nick Warner for this argument.}. This argument is similar to that ruling out Dirac strings, and in principle should also hold in the presence of D3 or anti-D3 branes.

A second possible criterion is requiring that the integrals of the $H_3$ from the origin to a certain holographic screen differ by an integer amount for two solutions in the same theory, or equivalently that the difference in the number of Seiberg duality cascades between two solutions dual to vacua of the same theory has to be integer-valued. This criterion has a clear physical justification for compact settings, where the KS throat is seen as the zoom-in of a compact CY, and where the three-cycle wrapped by $H_3$ that appears non-compact from a KS perspective is in fact embedded into a compact CY three-cycle. However, for a non-compact KS solution this criterion is very hard to justify from a holographic perspective, because it involves integrals over the whole bulk. 

One can also use the analysis of~\cite{Dymarsky:2005xt} to reverse-engineer a criterion that allows one to distinguish between vacua with various numbers of mobile branes without introducing any extra IR boundary conditions, and using only UV data. This third criterion (summarized in equation~\eqref{Y7Y1} for the first mesonic vacuum), if correct, certainly requires a more physical explanation.

Of course, another possibility is that the holography is just not refined-enough to distinguish between these different theories, especially because we are dealing with cascading solutions that are not asymptotically $AdS$, cannot be thought of as the near-horizon of any brane, and have an infinite charge unless one imposes an UV cut-off.

In this section, we will use the first criterion, and give a holographic recipe for distinguishing between asymptotically-KS vacua that have different numbers of mobile branes.

\subsection{Maxwell charge, Page charge and mobile D3-branes} \label{sec:charges}

For a supergravity solution with non-trivial Wess-Zumino terms one can generally 
define three different types of charges~\cite{Marolf:2000cb, Aharony:2009fc}, which we review in this section. The D3-Page charge, specialized to the KS background is
\bea
\cQ_{D3}^{Page}&=& \frac{1}{(4\, \pi^2)^2} \int _{T^{1,1}} \blp \cF_5-B_2\w F_3\brp\,.
\eea
This is conserved and is independent of the radius at which it is evaluated. In string theory it must also be quantized.
If we shift $B_2$ by a small gauge transformation $B_2\ra B_2+ d\Lam_1$ for some one-form $\Lam_1$, the charge stays invariant. In principle there are two independent ways to generate a non-zero, integer-valued $\cQ_{D3}^{Page}$ starting from the smooth KS background:
\bea
\cF_5&\ra& \cF_5 +  27 \, Q \pi \, \vol_{T^{1,1}} \, , \\
B_2&\ra& B_2 + \frac{p}{M}\, \pi \, \om_2 \, , \\
\Rightarrow \ \cQ^{Page}_{D3}&=&Q-p
\eea
where $(Q,p)\in \ZZ^2$, $M$ is related to $P$ by (\ref{PM}) and
\begin{align}
\vol_{T^{1,1}} =&\frac{1}{108} \, g_1 \wedge g_2 \wedge g_3 \wedge g_4 \wedge g_5 \  \nn \\
\omega_2 =& \frac{1}{2} (g_1 \wedge g_2 + g_3 \wedge g_4 ) \ .
\end{align}

Having $Q\neq 0$ generates a singularity in both the warp factor and $*\cF_5$, which one must interpret as due to $Q$ D3 branes smeared on the tip of the deformed conifold. On the other hand, the meaning of the singularity due to $p\neq 0$ is more subtle, and if one imposes as an IR regularity condition that the $B_2$ field at the KS tip be zero or an integer mod M, then $\cQ^{Page}_{D3}=Q$ measures the number (modulo $M$) of mobile BPS D3-branes in any particular KS background.   

The Maxwell D3-charge is 
\be
\cQ_{D3}^{Max}= \frac{1}{(4\pi^2)^2} \int_{T^{1,1}_{r_c}} \cF_5 \, ,
\ee
where the integral is performed on a Gaussian surface at the UV cut-off $r=r_c$.
There are two physically distinct contributions to the Maxwell charge, from mobile branes ($q_b$) and from charge dissolved in flux ($q_f$):
\bea
\cQ_{D3}^{Max}&=& q_b+q_f\,, \label{qsplit} \\
q_b&=&\frac{1}{(4\, \pi^2)^2}\, \int_{T^{1,1}_{0}} F_5 \,, \\
q_f&=&\frac{1}{(4\, \pi^2)^2}\, \Blp \int_{T^{1,1}_{r_c}}F_5- \int_{T^{1,1}_{0}} F_5\Brp =\frac{1}{(4\, \pi^2)^2}\, \int_{M_6} H_3\w F_3  \,.\label{qfdef}
\eea
The Maxwell charge depends on the scale at which it is measured, but if we fix a holographic screen, we expect physical processes to preserve its value at the screen. In particular, for a given scale, it must be the same if two solutions are to describe different vacua of the same theory. Using the Ansatz~\eqref{PTfluxes}, this is
\beq \label{QMaxPT}
\cQ_{D3}^{Max}=Q+\frac{4}{\pi}\, \left[(k-f)\, F +  2\, P\, f \right] \,   . 
\eeq
Note that if we set $\int_{S^2} B_2 = 0$ at the tip (i.e. requiring $f(\tau=0)=0$), then we have $Q=q_b=\cQ_{D3}^{Page}$ modulo $M$, while the second term in~\eqref{QMaxPT} gives the flux contribution to the Maxwell charge.

\subsection{A dictionary for the charges: two puzzles and two solutions} \label{sec:dictionary}

Our purpose is to establish using only UV data at a holographic screen whether two asymptotically-KS solutions describe vacua of the same theory. Any particular KS field theory is defined at a scale $\Lam_c$ through a gauge group $SU(N_1)\times SU(N_2)$ and the associated gauge couplings $(g_1,g_2)$.  The UV data of the supergravity theory consists of $\cQ^{Max}_{D5}(=M), \cQ^{Max}_{D3}, \int_{S^2} B_2 , \Phi$, and the ``standard lore'' dictionary between the supergravity UV data and the field theory is 
\bea \label{dictionary}
N_1&=& \cQ^{Max}_{D3} + \cQ^{Max}_{D5} \, , \label{lore1} \\
N_2&=& \cQ^{Max}_{D3} \, , \label{lore2}\\
\frac{4\, \pi^2}{g_1^2}+\frac{4\, \pi^2}{g_2^2}&=& \pi g_s^{-1}\, e^{-\Phi} \, , \label{lore3} \\
\Bslb \frac{4\, \pi^2}{g_1^2}-\frac{4\, \pi^2}{g_2^2}\Bsrb \, g_s\, e^{\Phi}&=& \left[ \frac{1}{2\, \pi \, \alpha'}\, \int_{S^2}B_2-\pi \right]\, \text{mod}\, (2\, \pi) \ , \label{couplingrunning}
\eea
as reviewed in~\cite{Herzog:2002ih}.
We can also trade the integral of $B_2$ for $\cQ^{Page}_{D3}$ using 
\be
\int_{S^2_{r_c}} B_2=(\cQ^{Max}_{D3} -\cQ^{Page}_{D3})/\cQ^{Max}_{D5} =  q_f/\cQ^{Max}_{D5} +\int_{S^2_{0}} B_2  \,.
\label{BPage}
\ee
As we will see shortly, this dictionary is in fact more involved.  

All this data is defined in the supergravity solution at some UV cut-off $r_c$ related to the field theory scale $\Lam_c$. To obtain this relation, we change to a radial coordinate $r$ such that the metric on the transverse six-dimensional space asymptotes to a warped conical metric: 
\be
ds_{10}^2=  h^{-1/2}\, ds_{1,3}^2+h^{1/2}\, ds_6^2 \, ,
\ee
with
\bea
ds_6^2 &\sim& dr^2 + r^2\, ds_{T^{1,1}}^2\,,\ \ \ r>>1\,. \non 
\eea
For any KS background \eq{KSbackground}, this $r$ coordinate is related to the deformed-conifold $\tau$ coordinate via
\bea
r^2&=&\frac{3}{2^{5/3}}\, \veps_0^{4/3}\, e^{2\tau/3} \label{rtau} \, .
\eea
The field theory cut-off $\Lam_c$ should then be identified with the holographic cut-off $r_c$. Note that from the point of view of the $\tau$ coordinate, the parameter $\veps$ only enters the function $A$ from the Ansatz, and changing it corresponds to a rescaling of the four-dimensional metric (see~\eqref{KSbackground}). 

We now run into the first puzzle, which can be expressed on the supergravity side alone. According to the dictionary above, since the field theory gauge group ranks depend only on $\cQ^{Max}_{D3}$ but not on $\cQ^{Page}_{D3}$ or $q_b$, one can see from equation (\ref{qsplit}) that the duals to solutions with different $q_b$ and $q_f$ but the same $\cQ^{Max}_{D3}$ have the same charges and should be dual to the same field theory. This is achieved by shortening the domain of integration in (\ref{qfdef}), which lowers $q_f$, and by increasing $q_b$ to compensate this. Hence, the only UV holographic data that will be different between, say, a solution with no mobile branes and a solution with one mobile brane will be the integral of $B_2$ on the $S^2$. However, this difference is not gauge-invariant, and if one does not impose any infrared boundary condition on $B_2$, we can see from (\ref{BPage}) that this value is arbitrary, and hence nothing in the UV will distinguish between a solution with one mobile brane and one with no mobile brane; we expect this to be incorrect.

One way to remedy this is to impose an IR boundary condition, namely that the integral of $B_2$ on the shrunken $S^2$ at the tip be gauge-equivalent to zero. If so, then two solutions with different numbers of mobile branes and different $q_f$ will have different $B$ fields in the UV, and will correspond to different theories. The only situation when the UV fields will be the same is when the number of mobile branes differs by multiples of $M$, when indeed we expect these solutions to correspond to different vacua of the same theory~\cite{Dymarsky:2005xt}. In the next subsection we will illustrate this in detail using our perturbation theory machinery. 

The second quandary has to do with the field theory interpretation of two solutions that have the same $\cQ^{Max}_{D3}$ but different numbers of mobile branes. If one is to take a holographic screen at $r_c$ and use the dictionary (\ref{lore1},\ref{lore2},\ref{lore3},\ref{couplingrunning}), a solution with $p<M$ mobile branes and one with none will be dual to two field theories that have the same ranks of the gauge group at the same cutoff, but differ only in the coupling constant. Furthermore, a solution that has $\cQ^{Max}_{D3}=M+1$ at a holographic screen at $r_c$ will have $\cQ^{Max}_{D3}=M$ at a holographic screen placed further down in the infrared; this would appear to imply that a theory with rank $SU(2M+1)\times SU(M+1)$ at some energy flows at lower energies to a theory with rank $SU(2M)\times SU(M)$, then $SU(2M-1)\times SU(M-1)$, which is definitely incorrect.

A partial solution to this puzzle is given by a comment in~\cite{Herzog:2002ih}, where it was noted that one cannot relate the UV supergravity data to field theory data at an arbitrary UV holographic screen. The dictionary (\ref{lore1},\ref{lore2},\ref{lore3},\ref{couplingrunning}) can only be used at special values of $r_c$, given by the requirement that from the infrared up to that scale the number of duality cascades is an integer, or alternatively, that the value of $q_f$ is a multiple of $M$. This is a stronger requirement than demanding that the ranks of the putative dual gauge groups are integer-valued. We will call for convenience the holographic screens at which one can define the dictionary ``K-screens.''

However, this cannot be the whole story. As we can see from equation~\eqref{BPage}, this restriction alongside the requirement that $B_2$ be zero at the tip imply that the value of the $B_2$ integral at the K-screen is a multiple of $M$, and hence the two field theory coupling constants will have the same values at any K-screen. Thus, at those screens (which are the only places where the field theory has an approximate Lagrangian description), the right-hand side of equation~\eqref{couplingrunning} is always equal to $\pi$, and the coupling constant of one of the gauge group always becomes infinite. Conversely, out of the set of possible field theory data defined at a scale $\Lambda_c$ via the 4 parameters $N_1,N_2,g_1$ and $g_2$, the KS supergravity solutions would only describe field theories that belong to a codimension-one subspace, and hence not the most generic field theory. 

In order to avoid the above-mentioned problems, equations~\eqref{lore3} and~\eqref{couplingrunning} should be used to obtain the values of the coupling constants as a function of the corresponding energy $\Lambda_c$. However, the ranks of the gauge groups given in equations~\eqref{lore1},\eqref{lore2} must be read from the K-screen right above it. Those equations then provide the ranks of the gauge groups both at the scale corresponding to $r_c$ and at the scale corresponding to the K-screen above. The ranks do not change when one changes the position of the holographic screen by decreasing $r_c$, unless one crosses another K-screen, which corresponds to a Seiberg duality in the dual theory.

One can also ask how can a holographist tell, using purely UV data, where the K-screen lies. The answer is given by~\eqref{couplingrunning} -- the screen is at the location above $r_c$ where the $B_2$ integral is gauge equivalent to zero. Hence, if the $B_2$ integral at the tip is zero, this dictionary gives a way to relate all 4 parameters of the field theory to the four parameters of the supergravity solution, using UV data alone.

\subsection{Baryonic and Mesonic Branches - a Perturbation-Theory Analysis } \label{sec:barmes}

When the ranks of the two gauge groups are  
\be
N_1=(k+1)\, M\,,\ \  N_2 = k\, M\,,\ \ k\in \ZZ
\ee
the theory has two classically disconnected supersymmetric moduli spaces, the baryonic and mesonic branches~\cite{Dymarsky:2005xt}. For more general $(N_1,N_2)$ the mesonic branch is supersymmetric while the baryonic branch is lifted. It is instructive to use the dictionary above together with the infrared boundary condition for $B_2$ to demonstrate in the supergravity perturbation theory framework we have developed that when they exist, both the baryonic and mesonic branches are indeed different vacua of the same theory.

As mentioned in section~\eqref{sec:charges}, if one imposes $\int_{S^2} B_2 =0$ modulo $M$ at the tip, then the function $f$ shoud go to zero at the origin. On the other hand, we have from~\eqref{phi5IR} in Appendix~\ref{app:IRUVexps} that
\beq \label{phi50}
\tilde \phi_5(\tau=0)=f(\tau=0)+\frac{\pi\, Q}{2\, M} = Y_7^{IR} \, ,
\eeq
where we have set $Y_6^{IR}=0$ since this mode diverges as $1/\tau^3$, and we have used the relation between $P$ and $M$ from~\eqref{PM}. This implies that in our perturbation theory
\beq \label{QY7}
\cQ^{Page}_{D3} =Q=\frac{2}{\pi}\, M\, Y_7^{IR} \, .
\eeq
Setting this equal to an integer multiple of $-M$, leads to\footnote{\label{foot:chargeconv}In our conventions the KS background has negative D3 charge.} 
\bea
\cQ^{Page}_{D3} &=&- \ell \, M \, , \\
\Rightarrow Y_7^{IR} &=& - \frac{\pi}{2}\, \ell \, . \label{Y7ell}
\eea
Physically this corresponds to adding $\ell \, M>0$ mobile D3-branes smeared on the tip of the KS solution and for each $\ell \in \ZZ$ this provides the bulk dual to the $\ell$-th mesonic branch. Let us note for later use that from~\eqref{phi4IR}, Appendix~\ref{app:IRUVexps}, we get that the warp factor at the tip is 
\beq \label{phi4Q}
\tau \, \tilde{\phi}_4(0)= - 6\, \frac{M}{h_0}\, \left(\frac{2}{3}\right)^{\frac13}\, Y_7^{IR}=\frac{3}{h_0}\, \left(\frac{2}{3}\right)^{\frac13}\, \pi \, |Q| \, .
 \eeq

To compare the Maxwell charges of the baryonic and mesonic branches, we must demand that they are defined at the same scale $\Lam_c$. To do so we must address the fact that the constant $\veps_0$ appearing in~\eqref{rtau} is not gauge invariant and can be set to one by rescaling the space-time coordinates $x_\mu$. As such one would normally fix the gauge and eliminate this constant. Indeed, $\veps_0$ is dimensionful and just serves to fix the units which may as well be set to unity. However the ratio between the value of $\veps_0$ in two different KS vacua, such as the mesonic and baryonic branches, is dimensionless and physically relevant. 

This is similar to the familiar domain wall solution from one $AdS$ vacuum to another. In either vacuum the $AdS$ radius sets the units in which all other dimensionful numbers are measured but the ratio of the two radii is related to the ratio of central charges and is physically meaningful.  Having said this, it is important to establish that in our Ansatz the rescaling of $x_\mu$ is done by the constant shift in $A$, given in the UV by
\beq
A=\frac13\, (\tphi_4-\tphi_1)= -\frac15 \, Y_{1}^{UV} + {\cal O}(1/\tau) \, , 
\eeq
where we have preemptively used the UV boundary conditions~\eqref{X3Y5} introduced below. 
So, allowing for just $Y_7$ and $Y_1$ to be non-zero, we can find the supergravity solution of the mesonic branch as a perturbation of the baryonic branch. Using~\eqref{QMaxPT} and~\eqref{KSbackground}, along with~\eqref{phi5UV}-\eqref{phi7UV}, we find that in our perturbation theory the zeroth- and first-order Maxwell charge at a particular radius $r_c>>1$, is\footnote{See footnote~\eqref{foot:chargeconv}.}
\be \label{Qmaxsol}
\cQ^{Max}_{D3} = - \frac{8\, P^2}{\pi}\, (\tau-1) + \frac{8\, P}{\pi}\, Y_7^{UV} + \cO \left( e^{-\tau/3} \right) \, .
\ee
Using an expansion of $\veps$
\be
\veps = \veps_0\, \blp 1 + \frac{\veps_1}{\veps_0} + \cO(Z^2)\brp \, ,
\ee
where $\veps_0$ denotes that of the baryonic branch, it is apparent that if we want to stay at a fixed $r_c$, then~\eqref{rtau} requires at first order
\be \label{deltaepsilon}
\delta \tau= - 2\, \frac{\veps_1}{\veps_0} \, .
\ee
Demanding that $\cQ^{Max}_{D3}$ at $r_c$ is equal for the baryonic and mesonic vacua, yields the relation
\bea
\frac{\veps_1}{\veps_0}= - \frac{ Y_7^{UV}}{2\, P} \label{epsY7} \, .
\eea
Using~\eqref{YUVYIR} and the fact that $X_a=0$, we have $Y_7^{UV}=Y_7^{IR}$. 
Then, referring to~\eqref{Y7ell}, we have
\be \label{lepsilon}
\frac{\veps_1}{\veps_0}= \frac{\ell\, \pi}{M} \ ,
\ee
which is the first-order approximation to the known result $\veps_{\ell}=\veps_0 \, e^{\ell \, \pi / M}$~\cite{Dymarsky:2005xt, Dymarsky:2011pm}.

Now, we can find the value of the other integration constant, $Y_1$. Using the way that $\veps$ enters into the PT Ansatz through $A$, equation~\eqref{KSbackground} and the UV expansions of Section~\eqref{sec:UVexp} for $A=(\tphi_4-\tphi_1)/3$ we get
\be
\frac{\veps_1}{\veps_0}=-\frac{3\, Y_1^{UV}}{10}\, . \label{epsY1}
\ee
Combining this with~\eq{epsY7} results in an expression for $Y_1$ in terms of $Y_7$:
\be
Y_1^{UV}=\frac{5}{3\, P}\, Y_7^{UV} \, . \label{Y7Y1}
\ee 

The relations obtained in this subsection can also be used to formulate the second and the third criteria for distinguishing between asymptotically-KS solutions.

\section{Finding the anti-D3 brane solution}
\label{antiD3solution}

We can now summarize the necessary ingredients for identifying the candidate supergravity solution describing the backreaction of anti-D3 branes. Firstly, we must eliminate unphysical IR singularities. For many modes this is entirely unambiguous, for other modes this can be somewhat subtle and as such we will discuss each mode as it arises. Secondly, we demand that the UV asymptotics are the same as for the original KS solution which we are perturbing around. 

In total, we have sixteen integration constants but the seven physical modes (dual to seven gauge invariant operators) account for just fourteen of these. In addition, one is accounted for by the zero energy condition~\eqref{ZEC}, which we use to eliminate $X_5$:
\beq \label{ZEC5}
X_5= \frac1P\, \left( X_2 - \frac23 \, X_3 \right) - \frac32 \, X_7 \ .
\eeq
The zero-energy condition is necessary to completely fix the reparameterization invariance of the radial coordinate (see~\cite{Gubser:2001eg} for a very explicit description of this). The final mode corresponds to the rescaling of $x_{\mu}$ and for reasons discussed above this is an important physical constant which is given again by~\eqref{epsY1}. It was pointed out in the revised version of~\cite{Dymarsky:2011pm} that the two vacua of the Klebanov-Strassler theory necessarily have different values of $\veps$. With our technology we are able to in fact compute the precise ratio of $\veps$ in the two different vacua.

The reader who is more interested in the end-process and in seeing or using our solution than in the boundary conditions we imposed to pick it out of the full parameter space of first-order deformations around the Klebanov-Strassler background can directly proceed to Section~\ref{sec:finalsolt}.

\subsection{IR boundary conditions}

We impose that the divergences in the IR for all the fields are zero, except for $\tphi_4$ and $\sqrt{{\cal F}_5^2}$, the warp factor and 5-form flux along the brane, which should go respectively like $1/\tau$ and $1/\tau^2$ due to the anti-D3-brane sources. The latter means that $\tphi_5$ should go to a constant. 

From the divergent term in $\tphi_8$ appearing in equation~\eqref{phi8IR} of Appendix~\ref{app:IRUVexps}, one finds the first relation among $X$'s and $Y$'s parameters that must be enforced:
\beq
X_8 = \frac19 \, \left( h_0 \, X_1 - 3 \, P \, X_6 \right) \ .
\eeq
From the divergent terms in $\tphi_2$ we get upon using~\eqref{ZEC5} that
\beq \label{Y2X6}
Y_2^{IR} = 0 \ , \quad X_6 = \frac{h_0 \, X_1 - 3 \, X_4}{6 \, P} \, .
\eeq
Out of the divergent terms in $\tphi_3$ we set (after using~\eqref{ZEC5} and~\eqref{Y2X6})
\beq \label{Y3X4}
Y_3^{IR} = 0 \ , \quad X_4=\frac23 \, h_0 \, X_1 \ .
\eeq
Note that the $\log \tau / \tau$ term is automatically zero once we take into account~\eqref{Y2X6}. Finally, the divergent
term in $\tphi_6$ requires
\beq
Y_6^{IR}=0 \, .
\eeq
Likewise, the other piece is zero upon using~\eqref{Y2X6},~\eqref{Y3X4}. 

In summary, out of requiring IR regularity in all fields apart from the warp factor, we have obtained the following relations
\beq \label{IRregularity}
Y_2^{IR}=Y_3^{IR}=Y_6^{IR}=0 \ , \quad X_4=\frac23 h_0  X_1 \ ,  \quad
X_6 = -\frac{h_0}{6P} X_1 \ ,  \quad X_8 = \frac16 h_0 X_1 \ .
\eeq
They are part of the relations that pick out of the full space of first order KS deformations the candidate solution describing the dual to a metastable state, taking into account the backreaction of anti-branes onto the zeroth order background. Let us move on and impose the remaining IR boundary conditions. 
  
We will now impose that there are $\bar N$ anti-D3 sources at the tip. The IR regularity conditions~\eqref{IRregularity} yields
\beq \label{phi4phi5}
\tphi_5 (0) =  Y_7^{IR} \ ,  \\
\eeq
as in the supersymmetric case described in Section~\ref{sec:barmes}, equation~\eqref{phi50}. We require $Q=\bar N$ (cf.~footnote~\ref{foot:chargeconv}), which results in
\beq \label{Y7N}
Y_7^{IR} = \frac{\pi}{8\, P}\, \bar N \ ,
\eeq 
where we have used~\eqref{QY7} and~\eqref{PM}. On the other hand, the warp factor is such that
\beq
\tau \, \tphi_4 (0) &=& 8\, \left(\frac{2}{3}\right)^{\frac13} \, \left( h_0 \,
  X_1 - \frac{3\, P}{h_0}\,  Y_7^{IR} \right) \ .
\eeq
It ensues from requiring this exhibits the expected behavior for regular 3-branes (given in~\eqref{phi4Q}) that
\beq \label{X1N}
 X_1 = \frac{3\, \pi}{4\, h_0^2}\, \bar N \ .
\eeq

Before moving on to discussing UV boundary conditions in the subsequent section, we note that inserting~\eqref{X1N} in~\eqref{probeD3force} leads to the following expression for the force exerted on a D3-brane probing this backreacted supersymmetry-breaking solution:
\beq\label{KKLMMTforce}
F_{D3}=\frac{8\, \pi}{h_0^2}\, \frac{2^{2/3}\, \bar N}{(\sinh 2\, \tau - 2\, \tau)^{2/3}} \ .
\eeq
This is precisely equal to the force on a probe anti-D3 brane exerted by $\bar N$ D3-branes that is computed in KKLMMT~\cite{Kachru:2003sx}. This provides further support that our IR boundary conditions are the right ones for anti-D3 branes. 

\subsection{UV boundary conditions}

As part of our UV boundary conditions, we impose the absence of non-normalizable modes (we will come back to discussing this point in section~\ref{sec:gauginomass}).
Requiring no divergent terms in $\tphi_3$, $\tphi_4$ as well as $\tphi_5$, $\tphi_6$ and $\tphi_7$ implies 
\beq \label{X3Y5}
Y_4^{UV} =0 \ , \qquad X_3 = 0 \ , \qquad Y_5^{UV} = 0 \ .
\eeq
Requiring no $e^{-\tau/3}\sim 1/r$ terms in $\tphi_2$, and using~\eqref{X3Y5} then determines 
\beq\label{X2X7}
X_7= 0 \ , \qquad 
X_2 = - \frac29 \, h_0\, X_1 \ .
\eeq
Besides, we do not want to turn on the non-normalizable mode that shifts the dilaton, which would correspond in the gauge theory to changing the sum of the coupling constants for the gauge group. Hence, we must enforce that
\beq \label{Y80}
Y_8^{UV}=0 \ .
\eeq

From~\eqref{X3Y5} and~\eqref{Y80}, we see that the Maxwell charge in the UV is the same as in Section~\ref{sec:barmes}, equation~\eqref{Qmaxsol}. We should demand that at a given bulk radial slice $r$, this is the same as the Maxwell charge for the supersymmetric vacuum, which is in the (first) mesonic branch and has $M - \bar N = 4\, P -\bar N$ D3-branes at the bottom. Keeping in mind that $\veps$ is allowed to differ in the two vacua, which using~\eqref{rtau} implies that the Maxwell charges have to be evaluated at different $\tau$, we require that\footnote{See Figure~\ref{plot:allQ} below.}
\bea
\cQ^{Max}_{D3} &=&- \frac{8\, P^2}{\pi}\, \left( \tau_0 + \delta \tau_{ms} - 1 \right) + \frac{8\, P}{\pi}\, Y_7^{UV}  \label{QMaxms}\\
&\stackrel{!}{=}& - \frac{8\, P^2}{\pi}\, \left( \tau_0 + \delta \tau_{1} - 1 \right) - 4 \, P + \bar{N} \, . \label{QMaxsusy}\ 
\eea
Here $\delta \tau_1$ corresponds to the cut-off associated to the first mesonic branch. It is given by
\beq
\delta \tau_1 = - \frac{\pi}{2\, P} \, ,
\eeq
where we have used\footnote{Recall that $P = \frac{1}{4}\, M\, \alpha^{\prime}$. For convenience we have fixed $\alpha^{\prime} = 1$ throughout.}  (\ref{deltaepsilon}) and (\ref{lepsilon}) for $\ell=1$. We therefore have
\beq
 \frac{16\, P^2}{\pi}\, \frac{\veps_{ms}}{\veps_0} + \frac{8\, P}{\pi}\, Y_7^{UV} = \bar N \, .
\eeq
 
Using~\eqref{epsY1} to relate the change in $\veps$ to $Y_1^{UV}$ leads to
\be\label{Y7Y1antiD3}
- \frac{8\, P^2}{\pi}\, \frac{3}{5}\, Y_1^{UV} + \frac{8\, P}{\pi}\, Y_7^{UV} - \bar{N}=0 \, .
\ee
Note that if $Y_7^{UV}$ were equal to $Y_7^{IR}$, the latter being given in~\eqref{Y7N}, it would ensue that $Y_1^{UV}=0$ and no change in $\veps$ would be necessary. 
However, consequent on inserting all our boundary conditions apart from the one associated to $Y_1$ in~\eqref{YUVYIR}, one finds
\beq
 \frac{8\, P}{\pi}\, Y_7^{UV} = \frac{8\, P}{\pi}\, 5.64178 \, Y_7^{IR} = 5.64178 \, \bar N \ .
 \eeq
The shift in $\veps$ can be tuned to cancel the difference in the first-order Maxwell charge ${\cal Q}^{Max}$ between the anti-D3 and the supersymmetric solution. 

\subsection{The perturbative solution for anti-D3 branes in KS }\label{sec:finalsolt}

In summary, from the IR and the UV boundary conditions, all the integration constants turn out to be expressed in terms of the number $\bar N$ of anti-D3's at the tip of the throat. As a reminder, $h_0 = h(\tau = 0)$ denotes the zeroth order warp factor of the Klebanov-Strassler solution~\eqref{hintegral} evaluated at the tip. Below we collect the outcome of the analysis from the previous two subsections:

\begin{alignat}{3} 
X_1 &=\frac{3\, \pi}{4\, h_0^2}\, \bar N \, , & \quad \quad 
Y_1^{UV} &= \frac{3.03804}{P^2}\, \bar{N} \, , & \quad \quad Y_1^{IR} &= \frac{4.33971}{P^2}\, \bar{N} \,,
\nn \\
X_2 &=-\frac{ \pi }{6\, h_0}\, \bar N  \, & Y_2^{IR}&=0\,, &Y_2^{UV} &= -\frac{1.48261}{P^2 }\, \bar{N} \,,\nn \\
X_3 &=0 \ ,&  Y_3^{IR}&=0 \ , &Y_3^{UV} &= \frac{8.40238}{P^2}\, \bar{N} \,,\nn \\
X_4 &=\frac{ \pi }{2\, h_0}\, \bar N \ ,&Y_4^{UV}&=0\,, & \label{fullboundary}\\
X_5 &=-\frac{ \pi }{6\, P\, h_0 }\, \bar N \ ,& Y_5^{UV}&=0\,, & Y_5^{IR}& = \frac{0.70514}{P}
\, \bar{N}\,, \nn\\
X_6 &=-\frac{\pi }{8\, P\, h_0}\, \bar N \ , & Y_6^{IR}&= 0\,, & Y_6^{UV} &= -\frac{4.08244}{P}\, \bar{N}\,,\nn \\
X_7 &=0 \ ,  &Y_7^{IR}&= \frac{\pi}{8\, P}\, \bar N  \,, &Y_7^{UV}& = \frac{2.21552}{P}\, \bar{N}\,, \nn \\
X_8 &=\frac{\pi}{8\, h_0}\, \bar N \ , & Y_8^{UV}&= 0 \,, & Y_8^{IR} &= \frac{0.234935}{P^2}\, \bar{N}\,.\nn
\end{alignat}
All the constants in the leftmost and middle columns, with the exception of $Y_{1}^{UV}$, have been obtained by directly imposing boundary conditions in either the IR or UV. From there on, $Y_1^{UV}$ was obtained from $Y_7^{UV}$ via~\eqref{Y7Y1antiD3}. Finally, the rightmost column was derived from the numerical integration which is tabulated in~\eqref{YUVYIR}. We have not computed the value of $Y_4^{IR}$ as it is more involved than the others and we do not need it, but in principle it can be done through numerical integration of the analytic solution~\eqref{phi4appendix}.

\begin{figure}[h]
\begin{center}
\includegraphics[scale=0.7]{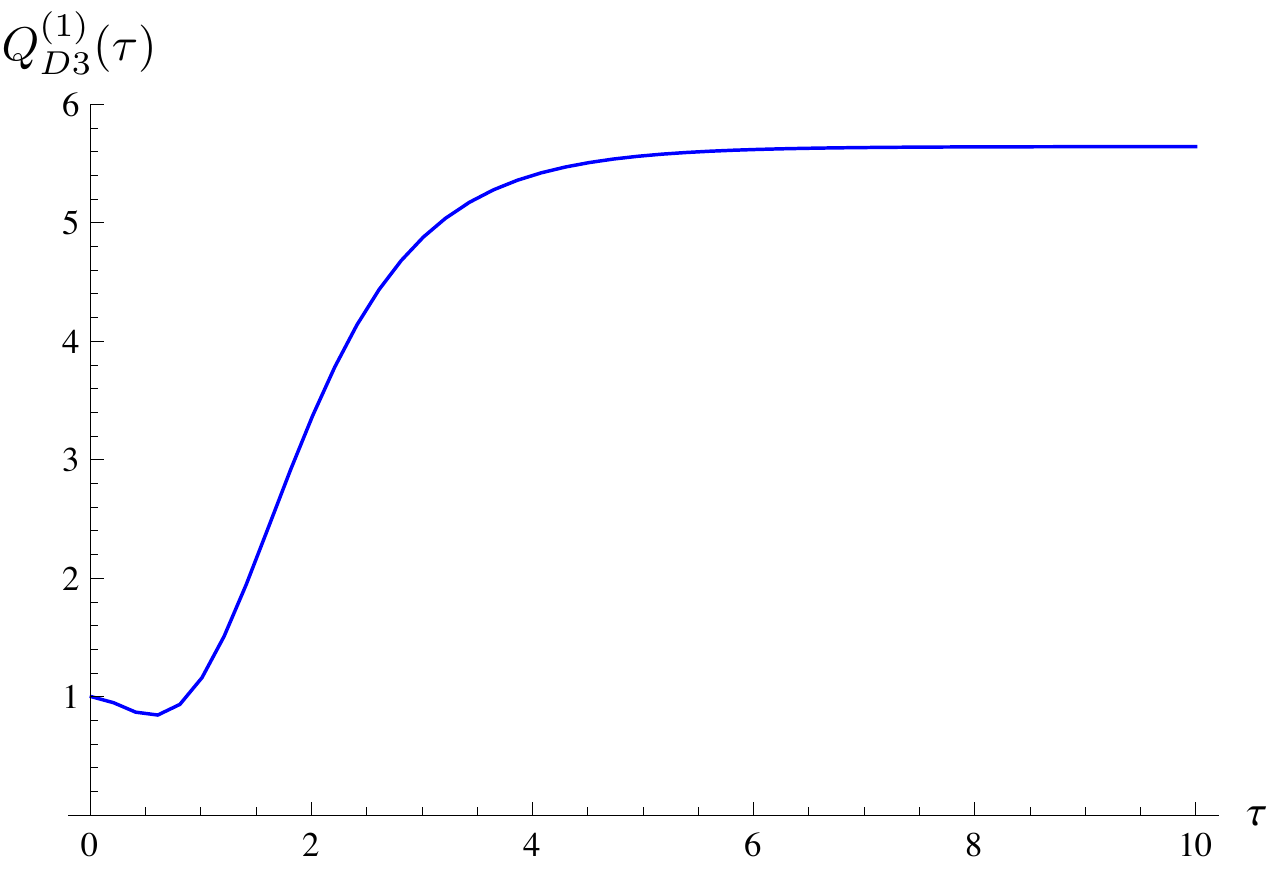} 
\caption{The profile of the first-order Maxwell charge for the anti-D3
  solution, setting $\bar{N}=1$.} \label{F5}
\end{center}
\end{figure}

It is interesting to observe the profile of the first-order perturbation to the Maxwell D3 charge ${\cal Q}^{Max}_{D3}$, given in Figure~\ref{F5} for $\bar N=1$ (see footnote~\ref{foot:chargeconv}). Note that it does not increase monotonically.

\begin{figure}[h]
\begin{center}
\includegraphics[scale=0.4]{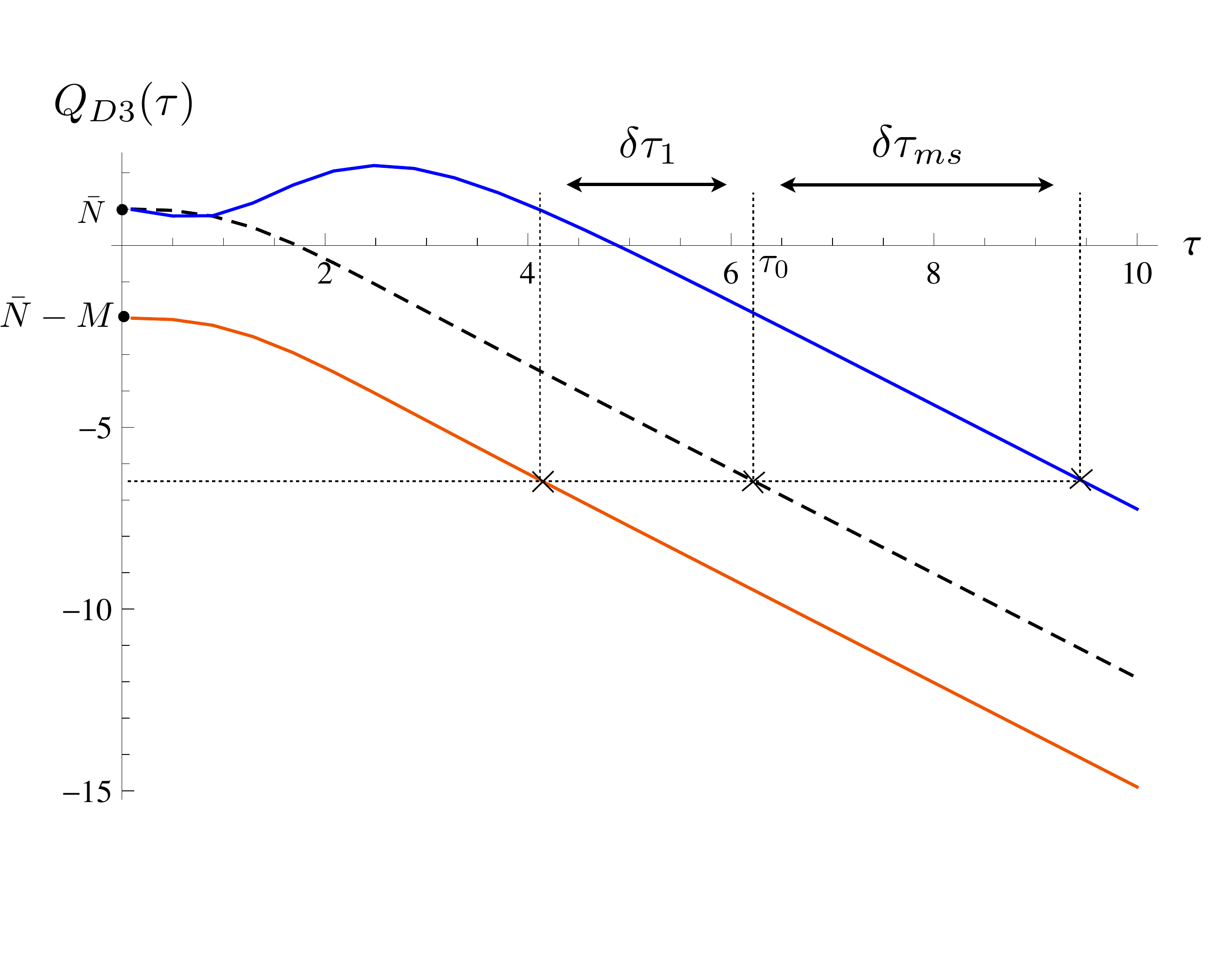} 
\vspace{-1cm}
\caption{Total Maxwell charge for the anti-D3 solution (blue), for the supersymmetric vacuum from the first mesonic branch (red) and for the ``would-be supersymmetric vacuum in the baryonic branch" (black dashed line), fixing $\bar{N}=1, M=3 \, (P=\tfrac34)$ .} \label{plot:allQ}
\end{center}
\end{figure}

On Figure~\ref{plot:allQ} we have plotted the total Maxwell D3 charge (i.e. the zeroth- plus first-order contributions) for the anti-D3-brane solution, alongside the Maxwell charge of the supersymmetric vacuum~\eqref{QMaxsusy}, the latter belonging to the first mesonic branch. For the purpose of illustrating equations~\eqref{QMaxms}-\eqref{QMaxsusy}, we also plot the ``would-be supersymmetric vacuum" in the baryonic branch, that we use as a reference to measure the difference in UV cut-off, $\delta \tau$. This branch obviously does not exist for ${\bar N}\neq 0$, but it is instructive to use it as yardstick.

 \newpage
 
\vspace{-1cm}


\subsection{Asymptotics of the solution}
 
The Green's function for the KS background diverges in the IR (\ref{hjIR}), and we denote the constant in its series expansion around $\tau = 0$ as $j_0$, Eq. (\ref{h0j0}). The IR and UV series expansions of the solution in terms of  $h_0$, $j_0$ and $X_1=\frac{3\, \pi}{4\, h_0^2}\, \bar N $ are as follows. 

\subsubsection{Behavior in the infrared}

In the IR the solution behaves as

\begin{align}
\tphi_8 & = 33.1634\, P^2 \, X_1 - \frac{512}{3}\, \left(\frac{2}{3}\right)^{2/3}\, P^2\, X_1\, \tau + \left[ \frac{64}{27}\, \left(\frac{2}{3}\right)^{1/3}\, h_0\, P^2\, X_1 + \frac{512}{27}\, \left(\frac{2}{3}\right)^{1/3}\, j_0\, P^2\, X_1 \right]\, \tau^2 \non\\
& + \mathcal{O}(\tau^3) \,
, \\
\tphi_2 & = -128 \, \left(\frac23\right)^{\frac23}\, P^2 \, X_1\, \tau + \frac{128}{81}\, \left(\frac23\right)^{\frac13}\, \big(
h_0 + 16\,P^2\, j_0 \big)  \, X_1\, \tau^2 + \mathcal{O}(\tau^3)  \, ,\\
\tphi_3 &= - \frac{224}{3}\, \left(\frac23\right)^{\frac23}\, P^2 \, X_1 \, \tau +
\frac{128}{405}\, \left(\frac23\right)^{\frac13}\, \big(h_0 + 136\,P^2\,j_0\big)
\, X_1 \, \tau^2 + \mathcal{O}(\tau^3) \, , \\
\tphi_1 & =   612.592\, P^2\, X_1 - \frac{704}{3}\, \left(\frac23\right)^{\frac23}\, P^2 \,
X_1\, \tau +
\frac{64}{405}\, \left(\frac23\right)^{\frac13}\, \big(7\, h_0 + 352\,P^2 \,j_0\big)
 \, X_1 \, \tau^2 + \mathcal{O}(\tau^3) \, , \\
\tphi_5 & = \frac{1}{6}\, h_0^2\, P\, X_1 - 4\, \left(\frac23\right)^{\frac13}\, h_0 \,P\, X_1
\, \tau^2 + \mathcal{O}(\tau^3) \, ,\\
\tphi_6 & = \frac{1}{6}\, h_0^2\, P\, X_1
-\frac{16}{3}\, \left(\frac23\right)^{\frac13}\,h_0\,P\,X_1 +
\frac{2}{81}\, \bigg(\frac{4\,h_0^2}{P} - 160\,h_0\,j_0\,P +
10451.6 P^3 \bigg)\, X_1\, \tau \nn \\
&\qquad + \bigg( \frac43\, \left(\frac23\right)^{\frac13}\,P\,h_0 -
  \frac{1280}{9}\, \left(\frac23\right)^{\frac23}\, P^3 \bigg)\, \,X_1\,\tau^2
+\mathcal{O}(\tau^3) \, , \\
\tphi_7 & = \frac{8}{3}\, \left(\frac23\right)^{\frac13}\, h_0\,P \,X_1\,\tau
- 83.769\, P^3\, X_1\, \tau^2 +\mathcal{O}(\tau^3) \,
,\\
\tphi_4
&=\bigg(4\, \left(\frac23\right)^{\frac13}\,h_0\,X_1\bigg)\, \frac{1}{\tau}
+ Y_4^{IR} + \bigg(
  \frac{8}{15}\, \left(\frac23\right)^{\frac13}\,h_0\,X_1 -\frac{64}{3}\, \left(\frac23\right)^{\frac23}\, P^2\,X_1 \bigg)\, \tau +
\mathcal{O}(\tau^2) \, ,
\end{align}

\subsubsection{UV behavior of the solution}

As for the ultra-violet behavior of the solution, it is described by the following UV series expansions:
\begin{align}
\tphi_8 & =-\frac{64}{3} \, 2^{1/3}\, e^{-4\tau/3}\, h_0\, X_1\, (\tau-1) - 288
\, 2^{2/3}\, e^{-8 \tau/3}\, P^2\, X_1  + \mathcal{O}(e^{-10\tau/3})\,
, \label{KSsolutionUV8}\\
\tphi_2& = - 418.571\, e^{-\tau}\, P^2\, X_1 + \frac{16}{3}\, 2^{1/3}\, e^{-7\tau/3}\,
h_0\, X_1 \, (1+8\tau) + \mathcal{O}(e^{-3\tau})\, ,\label{UV2}\\
\tphi_3 &=-\frac{32}{3} \, 2^{1/3}\, e^{-4\tau/3}\, h_0\, X_1 + 2\,
e^{-2\tau}\, \left( 1186.08 - 418.571\, \tau \right)\, P^2\, X_1 - \frac{1152}{5}\, 2^{2/3}\,
e^{-8\tau/3}\, P^2\, X_1 \non\\ &
+\mathcal{O}(e^{-10\tau/3})\, ,\\
\tphi_1 &=  428.85\, P^2\, X_1 + \frac83\,2^{1/3}\, e^{-4\tau/3}\, h_0\, X_1
-\frac23 \, e^{-2 \tau}\, \left( 1325.73 - 837.143\, \tau \right)\, P^2\, X_1
\nn \\
&\qquad + \frac{24}{5}\,2^{2/3}\, e^{-8\tau/3}\, P^2 \left( 29 + 40\, \tau \right)\, X_1
 +\mathcal{O}(e^{-10\tau/3})
\, ,\\
\tphi_5&= 312.743\, P^3\, X_1 + e^{-\tau}\, \left( -1361.84 + 418.571\, \tau \right)\, P^3\, X_1 - 4 \, 2^{1/3}\, e^{-4\tau /3}\, h_0\, P\, X_1\, (1 + 8\, \tau) \nn\\
&\qquad + 2\, e^{-2\tau}\, \left( 1361.84 - 837.143\, \tau \right)\, P^3\, X_1 + \mathcal{O}(e^{-7\tau/3}) \, ,\label{UV5}\\
\tphi_6 & = 312.743\, P^3\, X_1 + e^{-\tau}\, \left( 1361.84 - 418.571\, \tau \right)\, P^3\, X_1 - 4 \, 2^{1/3}\, e^{-4\tau /3}\, h_0\, P\, X_1 (1 + 8\, \tau) \nn\\
&\qquad + 2\, e^{-2\tau}\, \left( 1361.84 - 837.143\, \tau \right)\, P^3\, X_1 + \mathcal{O}(e^{-7\tau/3}) \, ,\label{UV6}\\
\tphi_7 & = e^{-\tau}\, \left( 943.269 - 418.571\, \tau \right)\, P^3\, X_1 \nn\\
&\qquad -\frac{4}{125} \, 2^{1/3}\, e^{-7\tau/3}\, h_0\, P\, \left(1199 +
80\, \tau\, (1+10\, \tau) \right)\, X_1 + \mathcal{O}(e^{-11\tau/3})\, , \label{UV7}\\
\tphi_4 &= 171.54 P^3 X_1+ \frac{4\,2^{1/3}\, e^{-4\tau/3}\, h_0\, (7 +
  32\, \tau )\, X_1}{3\, (4\, \tau-1)} - \frac{625.486 P^2 X_1}{ (4\, \tau-1)} +\mathcal{O}(e^{-2\tau})\, . \label{KSsolutionUV4}
\end{align}

\section{Additional Comments}
\label{oddsandends}

Having solved for the full space of linearized perturbations around
the Klebanov-Strassler background, we now discuss other solutions 
that we easily obtain as a by-product of our analysis, as well as other possible interpretations of our results.

\subsection{Relation to previous works}

The first attempt to construct the a linearized antibrane solution in the UV region alone was \cite{DeWolfe:2008zy}, which studied several of the $SU(2)\times SU(2)\times \ZZ_2$-invariant modes around the Klebanov--Tseytlin (KT) background~\cite{Klebanov:2000nc}. Since the KT solution is a subset of the parametrization~\eqref{PTmetric}--\eqref{PTfluxes} given by
\be \label{specKT}
y(\tau) = 0, \quad k(\tau) = f(\tau) , \quad F(\tau) = P \, ,
\ee
in our setup we can understand the perturbations around KTas
solutions of a reduced system of first-order differential equations in the
Borokhov--Gubser formalism. The details of this analysis, as
well as the relation with the notations of~\cite{DeWolfe:2008zy} can be found in
Appendix~\ref{app:KTperturbation}. The ``backreacted'' KT solution
contains some integration constants that cannot be fixed by infrared
boundary conditions, and hence we cannot relate them to the constant
$X_1$, which is proportional to $\bar N$.

We can directly compare the UV expansion of our full KS
solution~\eqref{KSsolutionUV8}-\eqref{KSsolutionUV4} to the
perturbed KT solution of~\cite{DeWolfe:2008zy} and we find the following crucial discrepancy: The correct UV expansion has terms of order $\cO(r^{-3})$ in (\ref{UV2},\ref{UV5},\ref{UV6},\ref{UV7}) while the first non-trivial terms in the solution of~\cite{DeWolfe:2008zy} are at $\cO(r^{-4})$. 

In hindsight this is not so surprising, since~\cite{DeWolfe:2008zy} only considered a subset of the modes, and furthermore, the KT solution precisely agrees with the UV limit of the KS solution only at leading order. At subleading order the KT solution has an ambiguity which can be fixed to agree with the UV limit of the KS solution but then the lower-order perturbation theory around each solution quantitatively differs. For this reason, we conclude that one cannot derive the correct UV expansion for the anti-brane solution by starting with the KT geometry. Another problematic issue with the Ansatz made in~\cite{DeWolfe:2008zy} is that, as we have explicitly demonstrated in this work, the anti-D3-branes turns on modes which are outside of the truncation, so it is not 
consistent to restrict oneself to this subset of mode. We refer the reader to Appendix~\ref{app:KTperturbation} for a thorough analysis of those issues.

\subsection{Gaugino masses} \label{sec:gauginomass}

As an additional outcome of our analysis, we can easily identify other
interesting solutions that correspond to different deformations of the dual
gauge theory. In particular, we can construct a solution in which the non-normalizable
UV modes $X_2$ and $X_7$ are turned on. They decay as $1/r$, and are associated 
to operators of dimension $\Delta = 3$, which correspond to deformations by gaugino mass
terms for each of the gauge groups, $\Tr (\lambda_1\lambda_1 \pm \lambda_2 \lambda_2)$. We will identify a one-parameter subfamily for which ${\cal Q}^{Max}_{D3}$
approaches the same constant value in the IR and in the UV, and therefore for
which the parameter $\veps$ does not need to be modified.

\begin{figure}[t]
\begin{center}
\includegraphics[scale=0.85]{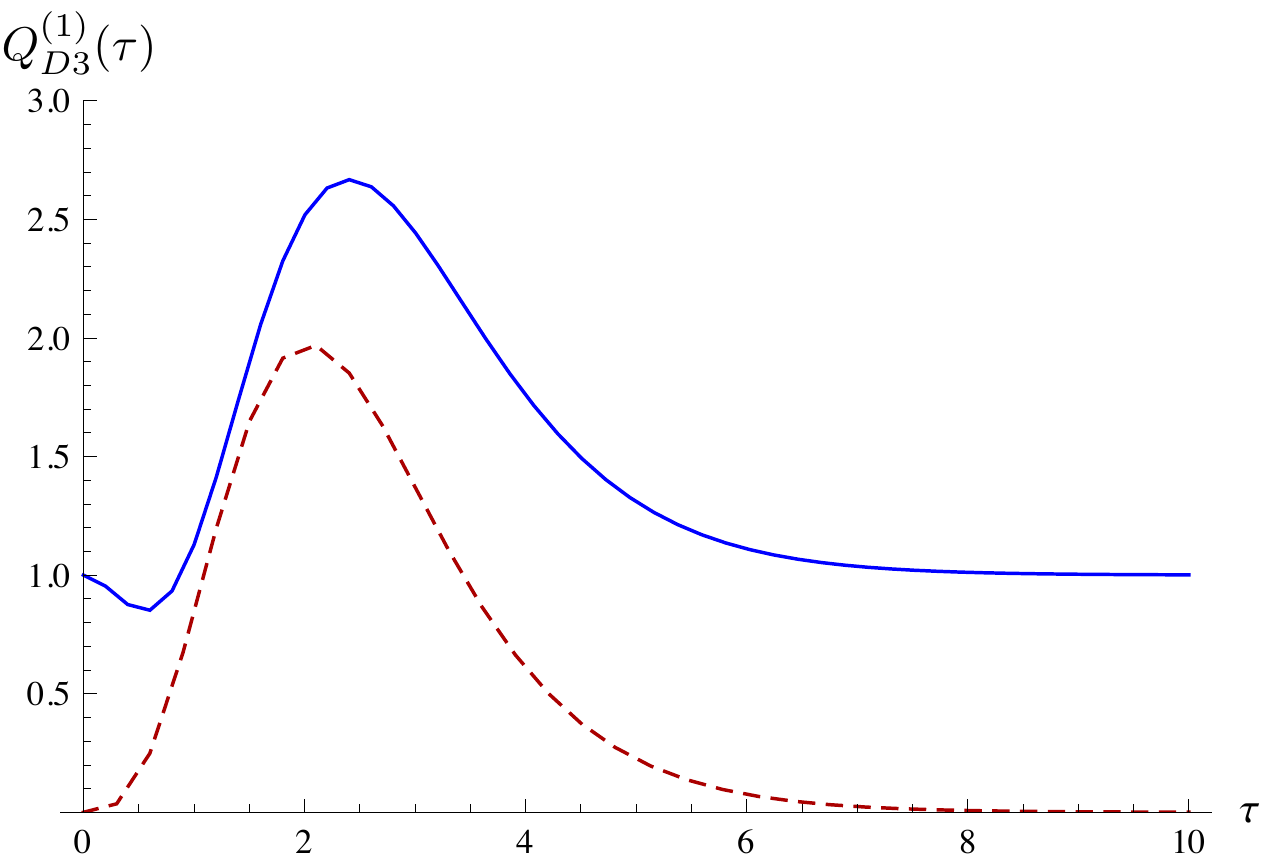} 
\caption{The profile of the first order Maxwell charge for the solution with gaugino masses turned on, satisfying the
  constraint~\eqref{relX2X7} (blue solid line). The plot is for
  $\bar{N} =1$ and $ X_7 = 1/(24 \, 2^{1/3} P^3) $. The red dashed curve is the profile
  for $\bar{N}=0$.} \label{phi5GM}
\end{center}
\end{figure}

The boundary conditions we have to impose are exactly the same as before, except that now we do not require~\eqref{X2X7}. 
Relaxing these, we find that the leading terms in the IR expansions are not modified,
and the value of $\tphi_5$ at the origin is still given
by~\eqref{phi4phi5}, together with the relations~\eqref{Y7N}),\eqref{X1N}
\be \label{phi5gaugino}
\tphi_5 (0) = Y_7^{IR} =\frac{h_0^2}{6P} X_1.
\ee
By using the UV/IR relation~\eqref{YUVYIR} we get that in the UV
\be
\tphi_5 (\infty) = Y_7^{UV} = 154.299\, P^3\,X_1 - 19.5477\,P\, (2X_2+P
X_7) \, .
\ee
Imposing $\tphi_5(0) = \tphi_5(\infty)$, we thus see that for the family of solutions
\be\label{relX2X7}
2X_2 + P X_7 = 5.05767 \,P^2 \,X_1
\ee
we get that the first order Maxwell D3 charge at infinity is the same as that of the supersymmetric vacuum 
with the same $\veps$ as for the
original KS background. The profile of the perturbation to the D3-brane Maxwell charge is shown in
Figure~\ref{phi5GM}, where it is plotted as a function of $\bar{N}$ using the condition from equation~\eqref{relX2X7}. 

We also note that by setting $X_1 = 0$, i.e. requiring that no anti-D3
brane be present at the origin, we obtain a family of
solutions parametrized by the constants $X_2$ and $X_7$ which in the
dual gauge theory describe soft supersymmetry breaking due to gaugino
mass terms. This solution encompasses the one built
in~\cite{Kuperstein:2003yt}, which corresponds\footnote{The constant $X$ in~\cite{Kuperstein:2003yt} is then related
  to $X_7$ by $X=-\frac12 X_7$ and their parameter $\mu$ is such that $\mu = 48\,2^{1/3}PX_7$.}  to the family $X_2 = P
X_7$ .

\subsection{Other UV boundary conditions}

In section \ref{antiD3solution} we have identified the anti-D3 backreacted solution using one of the three criteria to distinguish asymptotically-KS supersymmetric solutions that we have put forth in section \ref{section-distinguish}. The resulting solution has a different scale parameter $Y_1$ than its supersymmetric counterpart, and if the criterion that the NSNS $B_2$ field be zero at the KS tip is the correct one, then, putting aside concerns about the subleading singularity and about backreaction, the anti-D3 perturbative solution we have constructed describes a metastable stable of a supersymmetric KS field theory, and would be the first metastable solution constructed in supergravity. 

However, we can also ask whether this result holds if one imposes the other criteria, or if one insists, perhaps with a view towards embedding the KS solution in a compact setting, that the UV scale parameter $Y_1$ be the same as in the supersymmetric theory. It is not hard to see that if one imposes the criterion that the $H_3$ integral only jumps by integer units, one finds again that $Y_1^{UV}$ has to change; the anti-D3 solution is identical to the one we have written down above, and would be dual also to a metastable field theory vacuum.

If one on the other hand imposes the criterion that two vacua of the same theory must have a $Y_1^{UV}$ related to $Y_7^{UV}$ as in equation (\ref{Y7Y1}) (which also distinguishes between various supersymmetric KS vacua), or imposes the requirement that the UV scale must be the same as in the supersymmetric theory, then the resulting solution will have a different IR Maxwell charge than the one inferred from the UV data (essentially because antibranes give rise to negative charge dissolved in flux in their vicinity, as shown in Figure 4, and if one cannot make the throat longer to compensate for this, this charge will be visible at infinity). As a result, the relation between the force on a probe D3 brane and the anti-D3 charge of the background will not be the one of \cite{Kachru:2003sx}. If one then insists that this relation does not receive corrections at first order in the number of antibranes, as suggested by the no-screening results of \cite{Bena:2010ze}, then the anti-D3 solution must have a nontrivial $1/r$ mode turned on, of the type presented in the previous subsection, such that the contribution to the charge dissolved in flux from the antibranes is canceled by the contribution from the $X_2$ and $X_7$ modes. The value of this non-normalizable relevant perturbation can be easily read off from our analysis. Interestingly enough, such modes were argued in \cite{Baumann:2010sx} to be present when a KS solution is embedded in a stabilized flux compactification, and it is interesting to see if the relation between the anti-D3 charge and the strength of this mode that one can find here has any relevance  to this analysis.

\vspace{1cm}
 \noindent {\bf Acknowledgements}:
 \noindent We would like to thank D.~Baumann, D.~Diaconescu, A.~Dymarsky, C.~Herzog, I.~Klebanov, S.~Kuperstein, J.~Maldacena, L.~McAllister, M.~Petrini, H.~Verlinde, D.~Waldram and N.~Warner for useful discussions.
The work of G.~G.~and S.~M.~is supported by a Contrat de Formation par la Recherche of CEA/Saclay. The work of I.~B., M.~G.~and N.~H.~is supported by the DSM CEA/Saclay, the ANR grants 07--CEXC--006 and 08--JCJC--0001--0, and by the ERC Starting Independent Researcher Grants 240210 -- String--QCD--BH and 259133 -- ObservableString.

 \begin{appendix}
\section{First order deformation around a supersymmetric background} \label{app:BG}

\subsection{Our approach: from second order to twice as many first order differential equations}

The method introduced by Borokhov and Gubser~\cite{Borokhov:2002fm} to find the set of first order perturbations on top of a supersymmetric solution depending on a single radial variable $\tau$, relies on the existence of a superpotential $W$ whose square gives the potential obtained by reducing a supergravity Ansatz:
 \be
 V(\phi) = \frac{1}{8}\, G^{ab}\, \frac{\del W}{ \del \phi^a}\, \frac{\del W}{ \del \phi^b} \, .
 \ee
The fields $\phi^a$ ($a=1,...,n$) are expanded around their respective supersymmetric background values $\phi^a_0$,\beq
 \label{split}
 \phi^a = \phi^a_0 + \phi^a_1(Z) + {\cal O}(Z^2)\, ,
 \eeq
where $Z_A=(X_a,Y^a)$ represents the set of perturbation parameters and $\phi^a_1$ is linear in them.
The method amounts to splitting $n$ second-order equations into $2\, n$ first-order ones, out of which $n$ of them (those for the conjugate momenta $\xi_a$) form a closed set. The defining equations for the modes $\xi_a$ are
\beq
\label{xidef}
\xi_a \equiv G_{ab}(\phi_0)\, \left( \frac{d\phi_1^b}{d\tau} - M^b_{\ d}(\phi_0)\, \phi_1^d \right) \ , \qquad M^b{}_d\equiv\frac12 \, \frac{\partial}{\partial \fd}\, \left( G^{bc}\, \frac{\partial W}{\partial \fc} \right) \, .
\eeq
They measure the deviation from the BPS flow equations, i.e.~they are non-vanishing only for non-supersymmetric solutions. The set $(\xi_a,\phi^a)$
satisfies the equations:
\bea
\frac{d\xi_a}{d\tau} + \xi_b\, M^b{}_a(\fo) &=& 0 \, ,  \label{xieq} \\
\frac{d\fua}{d\tau} - M^a{}_b(\fo)\, \fub &=& G^{ab}\, \xi_b \label{phieq} \, ,
\eea
where~\eqref{phieq} is simply a rewriting of~\eqref{xidef} whereas the equations in~\eqref{xieq}
imply the equations of motion~\cite{Borokhov:2002fm}. Additionally, the functions $\xi_{a}$ should obey the zero-energy condition
\beq \label{ZECgen}
\xi_a\, \frac{d\fo^a}{d \tau} = 0 \, ,
\eeq
which stems from gauge-fixing the additional degree of freedom corresponding to reparametrisations of the radial variable (as explained very clearly in~\cite{Gubser:2001eg}). 

The $n$ integration constants arising upon solving~\eqref{xieq} are branded $X_a$, while those associated to~\eqref{phieq} are identified as $Y^a$.

\subsection{First-order equations for the deformations around KS}

Let us now review how the Borokhov-Gubser method is implemented for studying perturbations around the Klebanov-Strassler solution. There are eight functions in the   
Papadopoulos-Tseytlin Ansatz written in (\ref{PTmetric})-(\ref{PTfluxes}), $\phi^a=(x,y,p,A,f,k,F,\Phi)$, which is a consistent supersymmetric truncation of type IIB~\cite{Bena:2010pr,Cassani:2010na}.
Their zeroth-order values are available above in~\eqref{KSbackground}.

The field-space metric entering equation~\eqref{xidef} is computed out of the kinetic terms arising from the IIB reduction
\bea
\label{fieldmetric}
G_{ab}\, \phi^{\prime a}\, \phi^{\prime b} &=& e^{4\, p + 4\, A}\, \Big[ x'^2 + \frac12 \, y'^2 + 6\, p'^2 - 6\, A'^2 + \frac14 \, \Phi'^2 \non\\ && +
\frac14 \, e^{-\Phi-2\, x}\, \left( e^{-2\, y}\, f'^2 + e^{2\, y}\, k'^2 + 2\, e^{2\, \Phi}\, F'^2 \right) \Big]
\eea
The superpotential is found from the corresponding potential appearing from the reduction of the PT Ansatz:
\beq
\label{superpotential}
W(\phi)=e^{4\, A - 2\, p-2\, x} + e^{4\, A+4\, p}\, \cosh y + \frac12\, e^{4\, A+4\, p-2\, x}\, \left( f \, (2\, P-F) + k\, F \right) \, .
\eeq
In order to solve the system of equations~\eqref{xieq},\eqref{phieq} for the modes $\xi_a$ and $\phi_1^{a}$, we find it convenient to rotate to a different basis $(\tilde \xi_a, \tilde \phi^a)$, defined as follows in terms of the original fields:
\begin{align}
\txi_a& \equiv \left(3\, \xi_1 - \xi_3 + \xi_4, \, \xi_2, \, - 3 \, \xi_1 + 2\, \xi_3 - \xi_4, \, - 3 \, \xi_1 + \xi_3 - 2\, \xi_4, \, \xi_5+\xi_6, \, \xi_5-\xi_6, \, \xi_7, \, \xi_8 \right) \ , \\
\tphi^a&\equiv \left( x - 2\, p - 5\, A, \, y, \, x+3\,p, \, x-2\, p - 2\, A , \, f, \, k, \, F, \, \Phi \right) \ .
\end{align}
In the order we solve them, the system of first-order equations for the $\xi_a$~\eqref{xieq} reads 
\bea
 \txi_1'&=&e^{-2\, x_0} \, \left[ 2\, P\, f_0 - F_0\, \left( f_0 - k_0 \right) \right]\, \txi_1 \label{txi1eq} \, , \\
 \txi_4'&=& -e^{-2\, x_0}\, \left[ 2\, P\, f_0 - F_0\, \left( f_0 - k_0 \right) \right]\, \txi_1  \label{txi4eq} \, , \\
\txi_5'&=&-\frac13 \, P\, e^{-2\, x_0}\, \txi_1 \label{txi5eq} \, , \\
\txi_6'&=&-\txi_7-\frac13 \, e^{-2\, x_0} \, \left( P - F_0 \right)\, \txi_1  \label{txi6eq} \, , \\
\txi_7'&=&-\sinh(2\, y_0)\, \txi_5 - \cosh(2\, y_0)\, \txi_6 + \frac16 \, e^{-2\, x_0}\, \left( f_0-k_0 \right)\, \txi_1 \label{txi7eq} \, , \\
\txi_8'&=& \left( P \, e^{2\, y_0} - \sinh(2\, y_0)\, F_0 \right)\, \txi_5 + \left( P\, e^{2\, y_0} - \cosh(2\, y_0)\, F_0 \right)\, \txi_6 +\frac12 \, \left( f_0 - k_0 \right)\, \txi_7 \label{txi8eq} \, , \\
\txi_3'&=& 3\, e^{-2\, x_0 - 6\, p_0}\, \txi_3 + \left[ 5\, e^{-2\, x_0 - 6\, p_0} - e^{-2\, x_0}\,  \left( 2\, P\, f_0 - F_0\, \left( f_0-k_0 \right)\, \right) \right]\, \txi_1 \label{txi3eq} \, , \\
\txi_2'&=& \txi_2\, \cosh y_0 + \frac13 \, \sinh y_0 \, \left( 2\, \txi_1 + \txi_3 + \txi_4 \right) \non \, , \\
&& + 2\, \left[ \left( P\, e^{2\, y_0} - \cosh(2\, y_0)\, F_0 \right)\, \txi_5 + \left( P\, e^{2\, y_0} - \sinh(2\, y_0)\, F_0 \right)\, \txi_6 \right] \label{txi2eq} \, .
   \eea
Particularized to the deformation around KS, the system of $\phi_1^{a}$ equations is 
 \bea
  \tphi_8^{\prime} &=& - 4\, e^{-4\, A_0 - 4\, p_0}\, \txi_8 \label{phi8peq} \, , \\
  \tphi_2^{\prime} &=& - \cosh y_0\, \tphi_2 - 2\, e^{-4 \, A_0 - 4\, p_0}\, \txi_2 \label{phi2peq} \, , \\
 \tphi_3^{\prime} &=&  - 3\, e^{-6\,p_0-2\,x_0}\, \tphi_3 - \sinh y_0\, \tphi_2 - \frac16 \, e^{-4\, A_0 - 4\, p_0}\,  \left( 9\, \txi_1 + 5\, \txi_3 + 2\, \txi_4 \right) \, , \label{phi3peq}\\
  \tphi_1^{\prime} &=&  2\, e^{-6\, p_0-2\, x_0}\, \tphi_3 - \sinh y_0 \, \tphi_2 + \frac{1}{6} \, e^{-4 \, A_0 - 4\, p_0}\, \left( \txi_1 + 3\, \txi_4 \right)  \, , \label{phi1peq}\\
 \tphi_5^{\prime} &=&   e^{2\,y_0}\, \left( F_0 - 2\, P \right)\, \left( 2\, \tphi_2 + \tphi_8 \right) + e^{2\, y_0}\, \tphi_7
 - 2\, e^{-4 \, A_0 - 4\, p_0 + 2\, x_0 + 2\, y_0}\, \left( \txi_5 + \txi_6 \right) \, , \label{phi5peq}\\
 \tphi_6^{\prime}  &=&  e^{-2\,y_0}\, \left[ F_0 \, \left( 2 \, \tphi_2 - \tphi_8 \right) - \tphi_7 \right] - 2\, e^{-4\, A_0 - 4\,p_0 + 2\, x_0 - 2\, y_0}\, \left( \txi_5 - \txi_6 \right) \, , \label{phi6peq}\\
 \tphi_7^{\prime} &=& \half \, \Blp \tphi_5 - \tphi_6 + \left( k_0 - f_0 \right)\, \tphi_8 \Brp - 2 \, e^{-4\,A_0 -4\,p_0 + 2\, x_0}\, \txi_7 \, , \label{phi7peq} \\
  \tphi_4^{\prime} &=& \frac{1}{5}\, e^{-2\, x_0} \left[ f_0\, \left( 2\, P - F_0 \right) + k_0\, F_0 \right]\, \left( 2\, \tphi_1 - 2\, \tphi_3 - 5\, \tphi_4 \right) + \frac{1}{2}\, e^{-2\,x_0}\, \left( 2\, P - F_0 \right)\, \tphi_5 \non  \\
 && + \frac{1}{2}\, e^{-2\,x_0}\, F_0\, \tphi_6 + \frac{1}{2}\, e^{-2\,x_0}\, \left( k_0 - f_0 \right)\, \tphi_7 - \frac{1}{3}\, e^{- 4 \,A_0 - 4\, p_0}\, \txi_1 \, . \label{phi4peq}
 \eea

\section{The analytic solution space of deformations around KS} \label{app:solution}

Here we provide for handiness the solutions for the $\txi_a$'s and $\tphi^a$'s, in the order in which they were solved in our previous work~\cite{Bena:2011hz}. Of main interest are the $\tphi^a$ but we first had to solve for their ``conjugate momenta'' $\txi_a$ sourcing their equations.

\subsection{Analytic expressions for the $\xi_a$ modes}

\bea
\txi_1 &=& X_1 \, h(\tau) \, , \label{tX1}  \\
\txi_3 &= &\, - \frac{5}{3}\, X_1\, h(\tau) - \frac{32}{3}\, P^2\, X_1\, \text{csch}^2 \tau \, \left(\sinh \tau \, \cosh \tau - \tau \right)^{4/3} 
\non\\ 
&& - \frac{128}{9}\, P^2\, X_1\, \left( \sinh \tau \, \cosh \tau - \tau \right) \, j(\tau) + 2\, X_3\, \left(\cosh \tau \, \sinh \tau - \, \tau \right) \, , \label{tX3} \\
\txi_4 &=& - X_1 \, h(\tau) + X_4\, , \label{tX4} \\
\txi_5 &=& - \frac{16\, P}{3}\, X_1\, j(\tau) + X_5 \, ,\label{tX5} \\
\txi_6 &=& -\frac{1}{\sinh \tau}\, \lam_6(\tau) -\frac{\cosh \tau \, \sinh \tau - \tau}{2 \, \sinh \tau}\, \lam_7(\tau) \, , \label{txi6} \\
\txi_7 &=& -\frac{\cosh \tau}{\sinh^2 \tau}\, \lam_6(\tau) + \frac{-3+\cosh 2\, \tau + 2\, \tau\, \coth \tau}{4\, \sinh \tau}\, \lam_7(\tau)  \, , \label{txi7} \\
 \txi_8&=& P\, \left( \tau \, \coth \tau - 1 \right)\, \coth \tau \, \txi_5 - P\, \frac{\tau \, \coth \tau - 1}{\sinh \tau}\, \txi_6 - \frac{1}{6}\, X_1 \, h(\tau) + X_8 \, ,\label{tX8} \\
\txi_2 &= & - \frac{2}{3}\, X_3\, \tau \, \cosh \tau + \frac{1}{3}\, X_4\, \cosh \tau + P\, X_6\, \text{csch}\tau \, \left( \coth \tau - \tau \, \text{csch}^2 \tau \right) \non\\ 
&& + P\, X_5\, \text{csch} \tau \, \left( 1 - 2\, \tau \, \text{coth} \tau + \tau^2 \, \text{csch}^2 \tau \right) + X_2 \, \sinh \tau \non\\ 
&& + \frac{1}{2}\, P\, X_7\, \left( - 2\, \tau \, \coth^3 \tau + \text{csch}^2 \tau + \tau^2\, \text{csch}^4 \tau \right)\, \sinh \tau \non\\ 
&& - \frac{1}{108}\, X_1\, \bigg[ 3 \, \text{csch}^3 \tau \, h(\tau) \, \left( 6\, \tau - 5\, \sinh 2\, \tau + \sinh 4\, \tau \right) \non\\ 
&& + 2\, P^2\, \text{csch}^5 \tau \, \big( - 15 + 24\, \tau^2 + 16\, \cosh 2\, \tau - \cosh 4\, \tau - 32\, \tau \, \sinh 2\, \tau + 4\, \tau \, \sinh 4\, \tau \big)\non\\ 
&& \times \Big[ 4 \, \sinh^2 \tau \, j(\tau) - 6\, \left( \cosh \tau \, \sinh \tau - \tau \right)^{1/3} \Big]\, \bigg] \, ,
\eea
where
\bea
\lambda_6(\tau) &=& X_6 + \frac{1}{2}\, \left( - \tau + \coth \tau - \tau \, \coth^2 \tau \right)\, \txi_5(\tau) + \frac{1}{6}\, \frac{X_1}{P}\, h(\tau) \, , \\
\lambda_ 7(\tau) &=& X_7 - \csch^2\tau \, \txi_5(\tau) + \frac{16}{3}\, P\, X_1\, \text{csch}^2 \tau \, \left(\cosh \tau \, \sinh \tau - \tau \right)^{1/3} \non\\ && + \frac{64}{9}\, P\, X_1\, j(\tau) \,  .
\eea

\subsection{Analytic solutions for the $\phi_1^{a}$'s}

Holding our breath, we recap the analytic solutions for all eight $\tphi_1^a$ modes found in~\cite{Bena:2011hz}:

\begin{align}
\tphi_8 & = \, Y_8 - 64\, X_8\, j(\tau) + \frac{X_7}{P}\, h(\tau) - 64\, P\, X_6\, \int^{\tau} \frac{\left( u\, \coth u - 1 \right)}{\sinh^2 u \, \left( \cosh u \, \sinh u - u \right)^{2/3}} \, d u \non\\ & + \frac{2}{P}\, h(\tau)\, \txi_5(\tau) + \frac{16}{3}\, X_1\, \text{csch}^2 \tau \, \left( \cosh \tau \, \sinh \tau - \tau \right)^{1/3}\, h(\tau) + \frac{64}{9}\, X_1\, h(\tau) \, j(\tau)  \non\\ & 
+ \frac{64}{3}\, X_1\, \int^{\tau} \frac{\left( \sinh^2 u + 1 - u \,
    \coth u \right)}{\sinh^2 u \, \left( \cosh u \, \sinh u - u
  \right)^{2/3}}\, h(u)\, d u \, , \label{phi8appendix} \\[8pt]
\tphi_2 & =\text{csch} \tau \, \Lambda_2(\tau) \, , \label{phi2appendix}\\[8pt]
\tphi_3 &=\frac{1}{\sinh 2\, \tau - 2\, \tau}\, \Lambda_3 (\tau)  \label{phi3appendix}\,
,\\[8pt]
\tphi_1 &= \, Y_1 + \frac{40}{9}\, X_4\, j(\tau) - \frac{2}{3}\,
\tphi_3(\tau) - \frac{160}{9}\, X_3 \, \int^{\tau} \left( \cosh u \,
  \sinh u - u \right)^{1/3} \, d u \non\\ & + \frac{5}{3}\, \int \coth
u \, \Lambda_2^{\prime}(u)\, d u  -\frac{5}{3}\coth \tau \,
\Lambda_2(\tau)  + \frac{2560}{27}\, P^2\, X_1\, \int^{\tau}
\text{csch}^2 u \, \left( \cosh u \, \sinh u - u \right)^{2/3}\, d u
\non\\ & + \frac{10240}{81}\, P^2\, X_1\, \int^{\tau} \left( \cosh u
  \, \sinh u - u \right)^{1/3} \, j(u)\, d u  - \frac{80}{27}\, X_1\,
\int^{\tau} \frac{h(u)}{\left(\cosh u \, \sinh u - u \right)^{2/3}} \,
d u \, , \label{phi1appendix} \\[8pt]
\tphi_5 &= \frac{1}{2}\, \text{sech}^2(\tau/2)\, \left[ \tau + 2\,
  \tau \, \cosh \tau - \left( 2 + \cosh \tau \right)\, \sinh \tau
\right]\, \Lambda_5(\tau) + \frac{1}{1+\cosh \tau}\, \Lambda_6(\tau) +
\Lambda_7(\tau) \, , \label{phi5appendix} \\[8pt]
\tphi_6 &= \left[ \tau \, \left( 2 - \frac{1}{1 - \cosh \tau} \right)
  - \coth(\tau/2) + \sinh \tau \right]\, \Lambda_5(\tau) +
\frac{1}{1-\cosh \tau}\, \Lambda_6(\tau) + \Lambda_7(\tau) \, , \label{phi6appendix}
\\[8pt]
\tphi_7 &= \left( - \cosh \tau + \tau \, \text{csch} \tau \right)\,
\Lambda_5(\tau) - \text{csch}\tau \, \Lambda_6(\tau) \, , \label{phi7appendix} \\[8pt]
\tphi_4 & = \, \frac{1}{h(\tau)}\, \Big\{Y_4 - \frac{16}{3}\,
X_1\, \int^{\tau} \frac{h(u)^2}{\left( \cosh u \, \sinh u - u
  \right)^{2/3}} \, d u + 32\, P\, \int^{\tau} \frac{\left( u\, \coth
    u - 1 \right)\, \text{csch}^2 u \, \Lambda_6(u)}{\left( \cosh u
    \sinh u - u \right)^{2/3}} \, d u \non\\ & + 16\, P\, \int^{\tau}
\frac{\Lambda_7(u)}{\left(\cosh u \, \sinh u - u \right)^{2/3}} \, d u
+ \frac{32}{5}\, P\, \int^{\tau} \left( u\, \coth u - 1 \right)\,
\text{csch}^2 u \, \left( \cosh u \, \sinh u - u \right)^{1/3}\,
\non\\ & \times \left[ 5\, \Lambda_5(u) + 2\, P\, \left( - \tphi_1(u)
    + \tphi_3(u) \right) \right]\, du  \Big\} \, \label{phi4appendix} ,
\end{align}
where 

\begin{align}
\Lambda_2 & =  \, Y_2 - 16\, P\, X_7\, \int^{\tau} \frac{\left(- 2\, u \, \coth^3 u + \text{csch}^2 u + u^2\, \text{csch}^4 u \right)\, \sinh^2 u}{\left(\cosh u \, \sinh u - u \right)^{2/3}} \, d u \non\\ & - 32\, P\, X_6\, \int^{\tau} \frac{\coth u - u \, \text{csch}^2 u}{\left( \cosh u \, \sinh u - u \right)^{2/3}} \, d u - 32\, P\, X_5\, \int^{\tau} \frac{1- 2\, u\, \coth u + u^2\, \text{csch}^2 u}{\left( \cosh u \, \sinh u - u \right)^{2/3}} \, d u \non\\ & - \frac{32}{3}\, X_4\, \int^{\tau} \frac{\cosh u \, \sinh u}{\left( \cosh u \, \sinh u - u \right)^{2/3}} \, d u + \frac{64}{3}\, X_3\, \int^{\tau} \frac{u \, \cosh u \, \sinh u}{\left( \cosh u \, \sinh u - u \right)^{2/3}} \, d u \non\\ & - 48\, X_2\, \left( \cosh \tau \, \sinh \tau - \tau \right)^{1/3} + \frac{8}{9}\, X_1\, \int^{\tau} \frac{6\, u - 5\, \sinh 2\, u + \sinh 4\, u}{\sinh^2 u \, \left(\cosh u \, \sinh u - u \right)^{2/3}}\, h(u)\, d u \non\\ & 
- \frac{32}{9}\, P^2\, X_1\, \int^{\tau} \frac{- 15 + 24\, u^2 + 16\, \cosh 2\, u - \cosh 4\, u - 32\, u \, \sinh 2\, u + 4\, u \, \sinh 4\, u }{\sinh^4 u \, \left( \cosh u \, \sinh u - u \right)^{1/3}} \, d u \non\\ & 
+ \frac{64}{27}\, P^2\, X_1\, \int^{\tau} \frac{-15 + 24\, u^2 + 16\, \cosh 2\, u - \cosh 4\, u - 32\, u \, \sinh 2\, u + 4\, u \, \sinh 4\, u}{\sinh^2 u \, \left( \cosh u \, \sinh u - u \right)^{2/3}}\, j(u)\,  d u \, ,\\[8pt]
\Lambda_3 &=\, Y_3\, -\frac{32}{3}\, X_4\, \int^{\tau} \left( \cosh u
  \, \sinh u - u \right)^{1/3} \, du - \frac{112}{3}\, X_1\,
\int^{\tau} \left( \cosh u \, \sinh u - u \right)^{1/3}\, h(u)\, d u
\non\\ & - \frac{80}{3}\, \int^{\tau} \left( \cosh u \, \sinh u - u
\right)^{1/3}\, \txi_3(u) \, d u + 2\, \tau \, \coth \tau \,
\Lambda_2(\tau) - 2\, \int^{\tau} u\, \coth u \,
\Lambda_2^{\prime}(u)\, du \, , 
\end{align}
{\allowdisplaybreaks 
\begin{align}
\Lambda_5 &= \, Y_5 - \frac{1}{2}\, P\, \left(\tau \, \coth \tau -1 \right) \, \text{csch}^2 \tau \, \tphi_8(\tau) - 32\, P\, \int^{\tau} \frac{\left( u \, \coth u - 1\right)\, \text{csch}^2 u}{\left( \cosh u \, \sinh u - u \right)^{2/3}} \, \txi_8(u)\, d u \non\\ & +\frac{1}{4}\, X_7\, \int^{\tau} \text{csch}^4 u \, \left[ 2\, u \, \left( 2 + \cosh 2\, u \right) - 3\, \sinh 2\, u \right]\, h(u)\, d u -X_6\, \int^{\tau} \frac{2 + \cosh 2\, u}{\sinh^4 u}\, h(u)\, d u \non\\ & + \int^{\tau} \text{csch}^2 u \, \left[ - 3\, \coth u + u \, \left( 2 + 3\, \text{csch}^2 u \right) \right]\, h(u)\, \txi_5(u)\, d u - \frac{1}{2}\, P\, \frac{ \cosh \tau \,  \sinh \tau - \tau }{\sinh^4 \tau}\, \Lambda_2(\tau) \non\\ & + \frac{1}{2}\, P\, \int^{\tau} \text{csch}^4 u \, \left( \cosh u \, \sinh u - u \right)\, \Lambda_2^{\prime}(u)\, d u - \frac{X_1}{6\, P}\, \int^{\tau} \left( 2 + \cosh 2\, u \right)\, \text{csch}^4 u \, h^2(u)\, d u \non\\ & + \frac{16}{9}\, P\, X_1\, \int^{\tau} \text{csch}^4 u \, \left[ 2\, u \, \left( 2 + \cosh 2\, u \right) - 3\, \sinh 2\, u \right]\, j(u)\, h(u)\, d u \non\\ & + \frac{4}{3}\, P\, X_1\, \int^{\tau} \text{csch}^6 u \, \left( \cosh u \, \sinh u - u \right)^{1/3}\, \left[ 2\, u\, \left( 2 + \cosh 2\, u \right) - 3\, \sinh 2\, u \right]\, h(u)\, d u  \, ,   \\[8pt]
\Lambda_6 & = \, Y_6 - \frac{1}{2}\, P\, \left[ -\tau  + \coth \tau  +  \tau  \, \left( - 2 +  \tau  \, \coth  \tau  \right) \, \text{csch}^2  \tau  \right]\, \tphi_8( \tau ) 
\non\\&
- 32\, P\, \int^{\tau} \frac{\left[-  u  + \coth  u  +  u  \, \left( - 2 +  u  \, \coth  u  \right)\, \text{csch}^2  u  \right]}{\left(\cosh  u  \, \sinh  u  -  u  \right)^{2/3}}\, \txi_8( u )\, d u  
\non\\& 
+ \frac{1}{2}\, X_7\, \int^{\tau} \left[ \cosh 2\,  u  +\text{csch}^2  u  \, \left( 3 + 2\,  u ^2 - 6\,  u  \, \coth  u  + 3\,  u ^2\, \text{csch}^2  u  \right) \right]\, h( u ) \, d u  
\non\\&
+X_6\, \int^{\tau} \text{csch}^2  u  \, \left[ 3\, \coth  u  -  u  \, \left( 2 + 3\, \text{csch}^2  u  \right) \right]\, h( u )\, d u  
\non\\&
+\int^{\tau} \left[ 1 + \left( 3 + 2\,  u ^2 - 6\,  u  \, \coth  u  \right)\, \text{csch}^2  u  + 3\,  u ^2\, \text{csch}^4  u  \right]\, h( u )\, \txi_5( u )\, d u  
\non\\&
-\frac{1}{2}\, P\, \left[ 2\, \coth^2 \tau  \, \left( - 1 +  \tau  \, \coth  \tau  \right) + \text{csch}^2  \tau  -  \tau ^2\, \text{csch}^4  \tau  \right] \Lambda_2( \tau ) 
\non\\&
+ \frac{1}{2}\, P\, \int^{\tau} \left[ 2\, \coth^2  u  \, \left( - 1 +  u  \, \coth  u  \right) + \text{csch}^2  u  -  u ^2\, \text{csch}^4  u  \right]\, \Lambda_2^{\prime}( u )\, d u  
\non\\&
+ X_1\, \int^{\tau} \bigg\{ \frac{\text{csch}^4 u\, \left[ - 2\,  u  \left( 2 + \cosh 2\,  u  \right) + 3\, \sinh 2\,  u  \right]}{12\, P} h( u )\, + \frac{1}{36}\, P\, \text{csch}^6 u
\non\\&
\times \left[ 8 \, j(u)\, \sinh^2  u  + 6\, \left( \cosh  u  \, \sinh  u  -  u  \right)^{1/3} \right] \Big[ - 28 + 32\,  u ^2 + \left( 31 + 16\,  u ^2 \right)\, \cosh 2\,  u  \non\\ & - 4\, \cosh 4\,  u  + \cosh 6\,  u  - 48\,  u  \, \sinh 2\,  u  \Big] \bigg\} \, h(u)\, d u \\[8pt]
\Lambda_7 &= \, Y_7 + P\, \left[ -  \tau + \coth  \tau +  \tau \, \left( - 2 +  \tau \, \coth  \tau \right)\, \text{csch}^2  \tau \right] \, \tphi_8( \tau ) \non\\ & + 64\, P\, \int^{\tau} \frac{\left[ -  u + \coth  u +  u \, \left( - 2 +  u \, \coth  u \right)\, \text{csch}^2  u \right]}{\left( \cosh  u \, \sinh  u -  u \right)^{2/3} } \, \txi_8( u )\, d u \non\\ & + X_7\, \int^{\tau} \left[ - 1 + \left( - 3 - 2\,  u ^2 + 6\,  u \, \coth  u \right)\, \text{csch}^2  u - 3\,  u ^2\, \text{csch}^4  u \right]\, h( u ) \, d u \non\\ & + X_6\, \int^{\tau} \text{csch}^4  u \, \left[ 2\,  u \, \left( 2 + \cosh 2\,  u \right) - 3 \, \sinh 2\,  u \right] \, h( u )\, d u \non\\ & 
+ \int^{\tau} \left[ - 2 - 2\, \text{csch}^2  u \, \left( 3 + 2\,  u ^2 - 6\,  u \, \coth  u + 3\,  u ^2\, \text{csch}^2  u \right) \right]\, h( u )\, \txi_5( u )\, d u \non\\ &
- P\, \text{csch}^2 \tau \, \left( 1 - 2\,  \tau \, \coth  \tau +  \tau ^2\, \text{csch}^2  \tau \right)\, \Lambda_2( \tau ) \non\\ &
+ P\, \int^{\tau} \text{csch}^2 u \, \left( 1 - 2\,  u \, \coth  u +  u ^2\, \text{csch}^2  u \right)\, \Lambda_2^{\prime}( u )\, d u \non\\ & 
+ X_1\, \int^{\tau} \bigg\{ \frac{\text{csch}^4 u \, \left[ 2\,  u \, \left( 2 + \cosh 2\,  u \right) - 3\, \sinh 2\,  u \right]}{6\, P}\, h(u) - \frac{1}{9}\, P\, \text{csch}^6  u \, \non\\ & 
\times \left[ 8\, j(u) \,\sinh^2  u + 6\, \left( \cosh  u \, \sinh  u -  u \right)^{1/3} \right]\, \non\\ & 
\times \left[ - 9 + 16\,  u ^2 + 8\, \left( 1 +  u ^2 \right)\, \cosh
  2\,  u + \cosh 4\,  u - 24\,  u \, \sinh 2\,  u \right] \bigg\} \, \label{applambda7}
h(u)\, d u \, . 
\end{align} }

\section{IR and UV expansions of our analytic solutions} \label{app:IRUVexps}

\subsection{IR expansions}\label{subsecIRexpansions}

The IR behavior of the modes is obtained by Taylor expanding $h$, $j$
and the integrands in (\ref{phi8appendix}-\ref{phi4appendix}), performing the indefinite integral over $\tau$ (instead of the integral from 1 to $\tau$), and adding an integration constant $Y_a^{IR}$ (since the conjugate momenta $\xi_a$ do not involve integrals other than $h$ and $j$, we do not have to introduce a second set of integration constants $X^{IR}$ different from the one used in (\ref{tX1})-(\ref{tX8})). 

The IR expansions of $h$ and $j$ are given by
\begin{align} 
h_{IR} & = h_0 -\frac{16}{3}
  \left(\frac23\right)^{\frac13}P^2\tau^2 + \cO (\tau^3) \, , \nn \\
  j_{IR}&=-\frac{1}{\tau}\left(\frac{3}{2}\right)^{\frac23}+j_0-\frac{1}{5}
\left(\frac{2}{3}\right)^{\frac13} \tau+\cO (\tau^3) \, , \label{hjIR}
\end{align}
where
\begin{align} \label{h0j0}
h_0  = 18.2373P^2, \qquad j_0  = 0.836941 \ . 
\end{align}

In the order that those equations were solved and to the order of expansions that we need, the IR asymptotics of the $\tphi^a$ modes are given by
\begin{align}
\tphi_8&= \frac{1}{\tau} \frac{32}{3}
\left(\frac{2}{3}\right)^{\frac13} \left(-h_0 X_1+3 PX_6+9X_8\right)+
Y_8^{IR}+  \cO(\tau)\, , \label{phi8IR}  
\end{align}
\begin{align}
\tphi_2&=\frac{1}{\tau} Y_2^{IR} +\frac{\log \tau}{\tau} \left(\frac{16}{3} \left(\frac{2}{3}\right)^{\frac13} \left(h_0 X_1-3 (X_4+2 PX_6)\right) \right) +8 \left(\frac{2}{3}\right)^{\frac13} (-6X_2+4X_3+6 PX_5+9 PX_7) \nn \\
&+\cO(\tau) \, ,\label{phi2IR}  
\end{align}
\begin{align}
\tphi_3&=\frac{3Y_3^{IR}}{4 \tau^3} + \frac{1}{\tau} \left( \frac{Y_2^{IR}}{2}-\frac{3Y_3^{IR}}{20}+ \frac{4}{3} \left(\frac{2}{3}\right)^{\frac13} h_0 X_1+8 \left(\frac{2}{3}\right)^{\frac13} PX_6\right) \nn \\
&+ \frac{\log \tau}{\tau}
\left( \frac{8}{3} \left(\frac{2}{3}\right)^{\frac13} \left(h_0 X_1-3 (X_4+2 PX_6)\right) \right) 
  +\cO(\tau) \, ,\label{phi3IR} 
  \end{align}
\begin{align}
\tphi_1&=- \frac{1}{\tau^3} \frac{Y_3^{IR}}{2} + \frac{1}{\tau} \Blp -2Y_2^{IR} + \frac{Y_3^{IR}}{10}-\frac{4}{3} \left(\frac{2}{3}\right)^{\frac13} (4 h_0 X_1-3(5 X_4+12 P X_6)) \Brp \non \\
& + \frac{\log \, \tau}{\tau} \Blp -\frac{32}{3} \left(\frac{2}{3}\right)^{\frac13} \left(h_0 X_1-3 (X_4+2 PX_6)\right) \Brp + Y_1^{IR} \nn \\
&+ \log \tau \left(\frac{40}{3} \left(\frac{2}{3}\right)^{\frac13} (-6X_2+4X_3+6 PX_5+9 PX_7)\right) 
  +\cO(\tau) \, , \label{phi1IR} 
  \end{align}
\begin{align}
\tphi_5&= \frac{Y_6^{IR}}{2} + Y_7^{IR}   \nn \\
& +\tau^2 \left(-\frac{PY_2^{IR}}{2}-\frac{Y_6^{IR}}{8}+ \frac{1}{36\, P} h_0^2 X_1-4 \left(\frac{2}{3}\right)^{\frac13} PX_4+ \frac{1}{6} \left(-32\, 2^{\frac13} 3^{\frac23} P^2 +h_0\right) X_6-8\ 2^{\frac13} 3^{\frac23} PX_8\right) \nn\\
&+ \tau^2 \log \tau \left(   -\frac{8}{3}
  \left(\frac{2}{3}\right)^{\frac13} P \left(h_0 X_1-3 (X_4+2
    PX_6)\right)\right) +\cO(\tau^3) \, ,
\label{phi5IR} 
\end{align}
\begin{align}
\tphi_6&= \frac{1}{\tau^2} \Blp -2 Y_6^{IR} +\frac{8}{3} \left(\frac{1}{6 P} h_0^2 X_1+h_0 X_6\right) \Brp  \non \\
&+\Blp  \frac{Y_6^{IR}}{6} + Y_7^{IR} -\frac{2 PY_2^{IR}}{3}-\frac{128}{9} \left(\frac{2}{3}\right)^{\frac13} h_0 P X_1+\frac{2}{27P} h_0^2 X_1+16 \left(\frac{2}{3}\right)^{\frac13} PX_4  \nn \\
& +\left(-\frac{64}{3} \left(\frac{2}{3}\right)^{\frac13} P^2+\frac{4}{9} h_0 \right) X_6-32 \left(\frac{2}{3}\right)^{\frac13} PX_8 \nn
\Brp \\
&-\log \tau \Blp \frac{32}{9} \left(\frac{2}{3}\right)^{\frac13} P \left(h_0 X_1-3 (X_4+2 PX_6)\right)   \Brp +\cO(\tau) \, ,\label{phi6IR} 
\end{align}
\begin{align}
\tphi_7&= \frac{1}{\tau} \left( - Y_6^{IR} -\frac{2}{3} \left(\frac{1}{6P} h_0^2 X_1+h_0 X_6\right) \right) \nn \\
&+ 
\tau \left(\frac{P Y_2^{IR}}{3}+\frac{Y_6^{IR}}{6}
+\frac{64}{9} \left(\frac{2}{3}\right)^{\frac13} h_0 P X_1+\frac{1}{54P} h_0^2 X_1
-\frac{8}{3} \left(\frac{2}{3}\right)^{1/3} P X_4+\frac{1}{9} h_0  X_6-16 \left(\frac{2}{3}\right)^{\frac13} P X_8
 \right)  \nn
 \\
& +\tau \log \tau \left( \frac{16}{9} \left(\frac{2}{3}\right)^{\frac13} P \left(h_0  X_1-3 (X_4+2 P X_6)\right)\right) + \cO(\tau^2)\, , \label{phi7IR} \\
\tphi_4&=  \frac{1}{\tau} \Blp \frac{8}{9} \frac{P}{h_0} \left(\frac{2}{3}\right)^{\frac13} \left(-6 PY_3^{IR}-18Y_6^{IR}-27Y_7^{IR}+\frac{7}{P} h_0^2 X_1-12 h_0 X_6\right)
 \Brp  + Y_4^{IR} +\cO(\tau)\, . \label{phi4IR} 
\end{align}
Note that the constant term in $\tphi_2$ and the logarithmic term in $\tphi_1$ are identically vanishing once we impose the zero-energy condition (\ref{ZEC}). The relation between the constants $(X,Y^{IR})$ used here and those that first appeared in~\cite{Bena:2009xk}, which we denote $(\tilde{X}^{IR},\tilde{Y}^{IR})$, is summarized in the next subsection.

\subsubsection{Relation to the IR series expansion of~\cite{Bena:2009xk}}

The relation between the $X_a, Y_a^{IR}$ integration constants in this paper and the IR integration constants  in~\cite{Bena:2009xk}, which we call $\tilde X_a^{IR}, \tilde Y_a^{IR}$ depends $h_0$ and $j_0$, whose numeric values are given by~\eqref{h0j0}. We have
given by
\begin{eqnarray} \label{XIRtoX}
\tilde  X_1^{IR}&= h_0 X_1  \ , \quad  \quad \quad \quad \quad \quad \quad  & \tilde X_2^{IR}=\frac{1}{54} \left(-9 h_0+16 j_0 P^2\right) X_1 + \frac{1}{2} X_2+\frac{1}{6} X_4 \\
\tilde  X_3^{IR}&= -\frac{32 j_0 P^2}{9} X_1+ \frac{1}{2} X_3 \ , \quad  \quad   &\tilde X_4^{IR}= -h_0 X_1+ X_4 \nn \\
\tilde  X_5^{IR} &=-\frac{16 j_0 P}{3} X_1 +X_5 \ , \quad \quad  &\tilde  X_6^{IR} = \frac{1}{3} \left(-\frac{h_0}{P}+16 \left(2^{1/3} 3^{2/3}+j_0\right) P\right) X_1 -X_5 -2 X_6 \ , \nn \\
\tilde  X_7^{IR}& = -\frac{32 j_0 P}{9} X_1 -\frac12 X_7 \ , \quad &\tilde  X_8^{IR}=- \frac{h_0}{6}+\frac{8}{9} \left(2^{1/3} 3^{2/3}+2 j_0\right) P^2 X_1 -P X_5 -\frac{P}{2} X_7 + X_8 \nn
\end{eqnarray}
and
\bea \label{YIRtoY}
\tilde Y_a^{IR}&=&Y_a^{IR} \ \ \  \text{for} \,  a\neq 6,7 \ , \quad  \nn\\
\tilde Y_6^{IR}&=&  Y_6^{IR} + \frac{16 h_0}{3 P^2} \left(\frac{2}{3}\right)^{\frac13} X1 \ , \\
\tilde Y_7^{IR}&=&  Y_7^{IR} -\left(\frac{2^8}{3} \frac{P^2}{h_0} \left(\frac23\right)^{\frac23}-\frac83 \left(\frac23\right)^{\frac13} \right) h_0 \, X1  \  .   \nn
\eea

\subsection{UV expansions}  \label{sec:UVexp}

The UV asymptotics of $h(\tau)$ and $j(\tau)$ are
\begin{align} \label{hjUV}
h_{UV} & =  12\, 2^{1/3}P^2 (4\tau -1 )  e^{-4\tau/3} -\frac{128}{125}
\, 2^{1/3}P^2(12-85\tau+25\tau^2)e^{-10 \tau/3}+ \cO (e^{-16\tau/3}) \nn \\
  j_{UV}&= -\frac{3}{2^{2/3}} e^{-4\tau/3}
  -\frac{4}{25}\,2^{1/3}(3+10\tau)e^{-10\tau/3}+\cO  (e^{-16\tau/3})
  \, .
\end{align}
The UV expansions for the fields $\tphi_a$ are obtained by
performing an indefinite integration of the UV series of the
integrands as in the IR case. We call $Y_a^{UV}$ the 0th-order term in
the expansion for the field $\tphi_a$ (or $\Lambda_a$ if the former is
written as a product of the homogeneous solution times $\Lambda_a$)
\begin{align}
\tphi_8&=Y_8^{UV} + 12 \cdot 2^{1/3} \,e^{-4\tau/3} \Big( P (-1+ 4 \tau) (2
X_5 + X_7) + 8 X_8\Big) + \cO(e^{-8\tau/3}) \, , \label{phi8UV}  \\
\tphi_2&=-8 \cdot 2^{1/3} \, e^{-\tau/3} \Big(6 X_2 + (6 - 4 \tau) X_3 +
2 X_4 + 9 P X_7 - 6 P \tau X_7\Big) + 2\, e^{-\tau} Y_2^{UV} \nn \\
& +\cO(e^{-7 \tau/3})\, , \label{phi2UV}  \\
\tphi_3&= -5\cdot 2^{1/3} \,X_3 \, e^{2\tau/3}  -\frac43 \cdot 2^{1/3}
e^{-4\tau/3} \Big(108 X_2 + (336 - 137 \tau) X_3 + 48 X_4 \nn  \\
&- 108 P (-3 + \tau) X_7\Big)  +\cO(e^{-2\tau})\, , \label{phi3UV} \\
\tphi_1&= Y_1^{UV} -10\cdot 2^{1/3} \, X_3 \, e^{2\tau/3} + \frac23\cdot 2^{1/3}
e^{-4\tau/3} \Big(324 X_2 + (528 - 316 \tau) X_3 + 114 X_4 \nn \\
&+ 81 P (7 - 4\tau) X_7\Big) +\cO(e^{-2\tau}) \, , \label{phi1UV} \\
\tphi_5&= - \frac{Y_5^{UV}}{2}\, e^{\tau}  -Y_5^{UV} + Y_7^{UV} +
\tau (2 Y_5^{UV} -  P  Y_8^{UV})  \nn\\
&+6\cdot 2^{1/3} \, e^{-\tau/3} P
\Big(6 X_2 + (21 - 4 \tau) X_3 + 2 X_4 + 21 P X_7\Big) \nn \\
&+ \frac12 e^{-\tau} \Big((5 - 4 \tau) Y_5^{UV} + 4 Y_6^{UV} - 2 P
(-1 + 2 \tau) (Y_2^{UV} - Y_8^{UV})\Big) \nn \\
&+ 12 \cdot 2^{1/3} e^{-4\tau/3} P \Big(-12 (1 + \tau) X_2 - 15 X_3 - 4 X_4 + 
   2 \tau (X_3 + 4 \tau X_3 - 2 X_4 + 6 P X_5) \nn\\
&+ 3 P (-3 + \tau + 4 \tau^2) X_7 + 
   6 (P X_5 + X_8)\Big) +\cO(e^{-2\tau}) \, ,
\label{phi5UV} \end{align}
\begin{align}
\tphi_6&=  \frac{Y_5^{UV}}{2}\, e^{\tau} -Y_5^{UV}+ Y_7^{UV}   + \tau
 (2 Y_5^{UV} -  P Y_8^{UV}) \nn \\
&-6 \cdot 2^{1/3}\, e^{-\tau/3} P \Big(6 X_2 + (21 - 4 \tau) X_3 + 2
X_4 + 21 P X_7\Big) \nn \\
&+\frac12 e^{-\tau} \Big((-5 + 4 \tau) Y_5^{UV}-  4 Y_6^{UV} + 2 P
(-1 + 2 \tau) (Y_2^{UV} - Y_8^{UV})\Big) \nn \\
& +12 \cdot 2^{1/3} e^{-4\tau/3} P \Big(-12 (1 + \tau) X_2 - 15 X_3 - 4 X_4 + 
   2 \tau (X_3 + 4 \tau X_3 - 2 X_4 + 6 P X_5)  \nn \\
&+ 3 P (-3 + \tau + 4 \tau^2) X_7 + 
   6 (P X_5 + X_8)\Big) +\cO(e^{-2\tau})\, , \label{phi6UV} 
\end{align}
\begin{align}
\tphi_7&=  - \frac{Y_5^{UV}}{2}\, e^{\tau} + 18 \cdot 2^{1/3}
e^{-\tau/3} P \Big(-6 X_2 + (-9 + 4\tau) X_3 - 2 (X_4 + P (5 - 2 \tau)
X_7)\Big) \nn \\
& +e^{-\tau} \Big( (-\frac12 + 2 \tau) Y_5^{UV} - 2 Y_6^{UV} + P (Y_2^{UV} + 2 \tau Y_2^{UV} -
Y_8^{UV})\Big)  + \cO(e^{-7\tau/3}) \, ,\label{phi7UV} 
\end{align}
\begin{align}
\tphi_4&= \frac{Y_4^{UV}}{12\cdot
  2^{1/3}(4\tau-1)}e^{4\tau/3}-\frac{8\cdot 2^{1/3}(2\tau+1)\,
  X_3}{4\tau-1}e^{2\tau/3}   +\frac{2 Y_1^{UV}}{5} -  \frac{Y_5^{UV}}{P}
+\frac{Y_8^{UV}}{2} \nn\\
& - \frac{2 
  Y_7^{UV}}{P(4\tau-1) } + \frac{4\cdot 2^{2/3} (12 - 85\tau+25\tau^2)
  Y_4^{UV}}{1125(4\tau-1)^2}e^{-2\tau/3} +
\frac{2^{1/3} }{(4\tau-1)} e^{-4\tau/3}\Big( 18 (7 +8\tau) X_2 \nn \\
& + 32(2\tau +1) X_4 - 18P(7+8\tau)X_5 - 9 P(23+ 8\tau + 32\tau^2)X_7
-72 X_8 \nn\\
&+\frac{40803 - 170884 \tau + 161120 \tau^2 - 332800\tau^3)X_3}{375(4\tau-1)}\Big)   + \cO(e^{-2\tau}) \, . \label{phi4UV} 
\end{align}

\subsubsection{Relation to the UV series expansion of~\cite{Bena:2009xk}}

The relation between the $X_a, Y_a^{UV}$ integration constants used in the present paper and the UV integration constants introduced in~\cite{Bena:2009xk}, which we denote here $\tilde X_a^{UV}, \tilde Y_a^{UV}$, goes as follows:
\bea \label{XUVtoX}
\tilde  X_1^{UV}&=  -2^{\frac13} 4  P^2 X_1  \ ,    \quad \quad \quad  & \tilde X_2^{UV}=  \frac{1}{2}X_2 + \frac{1}{6} X_4 \, , \nn \\
 \tilde  X_3^{UV}&=\frac12 X_3  \ ,  \quad \quad \quad \quad \quad \quad  & \tilde X_4^{UV}=X_4 \, , \\
 \tilde  X_5^{UV}&= X_5 \ ,  \quad \quad \quad \quad  \quad \quad  & \tilde X_6^{UV}=-X_5 -2 X_6 \, ,\nn \\
\tilde  X_7^{UV}&= -\frac12 X_7  \ , \quad  \quad \quad \quad \quad  & \tilde X_8^{UV}=-PX_5 -\frac{P}{2} X_7 + X_8
\eea
and 
\bea \label{YUVtoY}
\tilde Y_a^{UV}&=&Y_a^{UV} \ \ \  \text{for} \,  a\neq 6 \ ,   \\
\tilde{Y}_6^{UV} &=& Y_6^{UV} - Y_2^{UV}+\frac12 Y_8^{UV}  \ . \nn \\
\eea

\section{The Klebanov--Tseytlin perturbation} \label{app:KTperturbation}

In our parametrization of the metric and the
fluxes~\eqref{PTmetric}--\eqref{PTfluxes}, the Klebanov-Tseytlin
background corresponds to the subset defined via~\eqref{specKT}.
At zeroth-order the fields $\phi_{KT}^a$ obey the flow equations 
\be
\frac{d \phi_{KT}^a}{d\tau} = \frac12 G^{ab}\frac{\partial W}{\partial \phi_{KT}^{a}} \, ,
\ee 
with the superpotential
\be
W(\phi)=e^{4\, A - 2\, p-2\, x} + e^{4\, A+4\, p}\, \left( 1 + P \, e^{-2\, x}\, f\right)\, .
\ee
These equations are solved by 
\begin{align}
A_0  &= -\frac14\log \left( h_{KT}(r)\right) \, , \nn\\
x_0 &= \frac12 \log \left( \frac{h_{KT}(r) r^4}{32\cdot
    2^{1/3}}\right)\label{appsolKT} \, , \\
p_0&=\frac16 \log \left(\frac{48\cdot
    2^{1/3}}{k_{KT}(r)r^4}\right) \, , \nn\\
f_0 &= P (1-3\log r) \, , \nn\\
\Phi_0 &=0 \, , \nn
\end{align}
where the warp factor $h_{KT}$ of the Klebanov-Tseytlin solution takes the following expression:
\be
h_{KT} = \frac{12\cdot 2^{1/3} P^2 (12 \log r -1)}{r^4} \, .
\ee

In order to match the UV asymptotic of the Klebanov--Strassler
modes~\eqref{KSbackground} we should use the relation $r=e^{t/3}$, while the perturbation
in~\cite{DeWolfe:2008zy} corresponds to changing the origin of the
$\log$ as follow: $\log r \rightarrow
\log r -\frac13$. This is equivalent to changing $\veps$. The relation
to the functions $a(r), b(r), k(r)$ used in~\cite{DeWolfe:2008zy} is the
following
\begin{align}
a(r) &= -\frac12(x(r) +\log 6) \, ,\\
b(r) &= -3p(r) -x(r) -\log 6 + \log (3\sqrt{6})\, ,\\
k(r)&= 6 f(r) \, ,
\end{align}
whereas the constant $\bar M$ is related to our P as
\be
\bar M = -18\, P \, .
\ee
If we consider linearized deformations around~\eqref{appsolKT}, it is quite simple
to solve analytically the linearized
equations~\eqref{xieq}--\eqref{phieq} for the five $\xi_a$ and five $\phi^a$ modes. In this way we get a
solution which contains terms up to the order $r^{-8}$. Henceforth, as a
bonus, we obtain the perturbation around the KT background that includes the mode responsible for the force on a probe
D3 brane discussed (but not worked out quantitavely) in~\cite{DeWolfe:2008zy}. The
results for the UV modes expansions are as follows:
is the following
\begin{align}
 \tphi_{\Phi} &= \frac{-288\cdot 2^{2/3}P^2 X_1}{r^8} + \frac{72
    \cdot 2^{1/3}  P X_f (1 + 4\log r)}{r^4}+ \frac{
   96 \cdot 2^{1/3} X_{\Phi}}{r^4} + Y_{\Phi} \, , \label{appKTphi8}\\
 \tphi_3 &= -\frac{1152\cdot 2^{2/3}P^2 X_1}{5 r^8}-10 \cdot 2^{1/3} r^2 X_3
-\frac{16 \cdot 2^{1/3} X_4}{r^4} + \frac{Y_3}{r^6} \, ,\\
  \tphi_1 &= \frac{24 \cdot 2^{2/3}P^2 X_1 (29 + 120\log r)}{5r^8} - 20 \cdot 2^{1/3} X_3  r^2
+ \frac{4 \cdot 2^{1/3} X_4}{r^4}-\frac{2 Y_3}{3 r^6} + Y_1 \, ,\\
 \tphi_{f} &=\frac{72 \cdot 2^{2/3}P^3 X_1 (12 \log r -1)}{r^8} + \frac{144\cdot
  2^{1/3} P^2 X_f (1 + 3\log r)}{r^4}\nonumber \\ 
&  + \frac{72 \cdot 2^{1/3}P
  X_{\Phi}}{r^4} - 3 P Y_8 \log r + Y_f \, ,\\
  \tphi_4&= \frac{48\cdot 2^{2/3} P^2 X_1 (-67  - 72\,\log r\,
 +2880\, \log^2r)}{25 r^8 (12 \log r-1)} - \frac{18 \cdot 2^{1/3} \,
  P\, X_f
    (11 + 24 \log r)}{r^4 (12 \log r -1)} \nonumber \\
& \quad+ \frac{72 \cdot 2^{1/3} X_{\Phi}}{r^4 (1- 12  \log r)} -\frac{16 \cdot
  2^{1/3} r^2 X_3 (1 + 6 \log r)}{12 \log r -1} + \frac{2 \cdot
  2^{1/3} X_4 (24 \log r - 5)}{r^4 (12 \log r -1)}  \nonumber\\
& \quad+ \frac{2Y_1}{5} -\frac{2 Y_f+ P r^4 Y_4}{P (12\log r -1)}- \frac{8 Y_3
  (30\log r - 7)}{75 r^6 (12 \log r -1)} - \frac{Y_{\Phi} (3 + 12 \log
  r)}{2 - 24 \log r}\, . \label{appKTphi4}
\end{align}
where $\tilde \phi_1, \tilde \phi_3, \tilde \phi_4$ are defined as in (\ref{tphidef}), and $\tilde \phi_f$ and $\tilde \phi_{\Phi}$ are respectively
the perturbations to the function $f$ (=$k$) and the dilaton. 

In order to compare to the full KS
solution~\eqref{KSsolutionUV8}--\eqref{KSsolutionUV4} we should
identify
\be\label{appKTtoKS}
X_f = X_5, \qquad X_{\Phi} = X_8 - P\, X_5,
\ee
and indeed after replacing $X_4$, $X_5$ and $X_8$ with the boundary
conditions given in~\eqref{fullboundary} the $r^{-4}$ and $r^{-8}$
terms agree.

By rescaling the radial coordinate $r$ we can compare to~\cite{DeWolfe:2008zy}. We get the following relation between their parameters ${\cal S}$, $\phi$, and our $X_f$, $X_{\Phi}$
\beq
{\cal S}= -96 \, 2^{1/3} P X_f \ , \qquad \phi= 24 \,2^{1/3} (7 P X_f
+ 4 X_{\Phi}) \ .
\eeq
Note that the IR boundary conditions relate $X_f$ and $X_{\Phi}$ to
$\bar N$
\begin{equation}
X_f=-\frac{1}{6 P h_0} \pi \bar N \ ,  \qquad X_{\Phi}=\frac{7}{24
  h_0} \pi \bar N\, .
\end{equation}
As a result, those parameters cannot be taken as independent ones,
contrary to what has been done in the literature.
By using these relations we see that 
\beq
\phi=0 \ .
\eeq
Note that this condition can be obtained by imposing just IR regularity conditions, and therefore any solution with 
a non-zero $\phi$ is singular in the IR. Imposing all those conditions, the $1/r^4$ terms agree with those
of~\cite{DeWolfe:2008zy}. However, the agreement is not complete: there is some discrepancy with the non-logarithmic term in $\tphi_f$.

\end{appendix}

\providecommand{\href}[2]{#2}\begingroup\raggedright\endgroup

\end{document}